\documentclass[aps,amssymb,amsmath,notitlepage,superscriptaddress,twocolumn,nofootinbib]{revtex4-1}

\usepackage{amsfonts}
\usepackage{graphicx,graphics,epsfig,times,bm,bbm,mathrsfs} 
\usepackage[normalem]{ulem}
\usepackage{subfigure}
\usepackage[pdfstartview=FitH]{hyperref}
\hypersetup{
    colorlinks=true,       
    linkcolor=cyan,          
    citecolor=magenta,        
    filecolor=magenta,      
    urlcolor=cyan,           
    runcolor=cyan
}
\usepackage[pdftex]{color}
\usepackage{upgreek}

\newcommand{\beq}{\begin{equation}}
\newcommand{\eeq}{\end{equation}}
\newcommand{\beqnn}{\begin{equation*}}
\newcommand{\eeqnn}{\end{equation*}}
\newcommand{\beann}{\begin{eqnarray*}}
\newcommand{\eeann}{\end{eqnarray*}}

\newcommand{\mc}{\mathcal}

\newcommand{\bes} {\begin{subequations}}
\newcommand{\ees} {\end{subequations}}
\newcommand{\bea} {\begin{eqnarray}}
\newcommand{\eea} {\end{eqnarray}}

\newcommand{\ignore}[1]{}

\begin{document}

\title{Scalable effective temperature reduction for quantum annealers\\ via nested quantum annealing correction}

\author{Walter Vinci}
\affiliation{Department of Electrical Engineering, University of Southern California, Los Angeles, California 90089, USA}
\affiliation{Department of Physics and Astronomy, University of Southern California, Los Angeles, California 90089, USA}
\affiliation{Center for Quantum Information Science \& Technology, University of Southern California, Los Angeles, California 90089, USA}
\author{Daniel A. Lidar}
\affiliation{Department of Electrical Engineering, University of Southern California, Los Angeles, California 90089, USA}
\affiliation{Department of Physics and Astronomy, University of Southern California, Los Angeles, California 90089, USA}
\affiliation{Center for Quantum Information Science \& Technology, University of Southern California, Los Angeles, California 90089, USA}
\affiliation{Department of Chemistry, University of Southern California, Los Angeles, California 90089, USA}

\begin{abstract}
Nested quantum annealing correction (NQAC) is an error correcting scheme for quantum annealing that allows for the encoding of a logical qubit into an arbitrarily large number of physical qubits. The encoding replaces each logical qubit by a complete graph of degree $C$. The nesting level $C$ represents the distance of the error-correcting code and controls the amount of protection against thermal and control errors. Theoretical mean-field analyses and empirical data obtained with a D-Wave Two quantum annealer (supporting up to $512$ qubits) showed that NQAC has the potential to achieve a scalable effective temperature reduction, $T_{\rm eff} \sim C^{-\eta}$, with $\eta \leq 2$. We confirm that this scaling is preserved when NQAC is tested on a D-Wave 2000Q device (supporting up to $2048$ qubits). In addition, we show that NQAC can be also used in sampling problems to lower the effective temperature of a quantum annealer. Such effective temperature reduction is relevant for machine-learning applications. Since we demonstrate that NQAC achieves error correction via an effective reduction of the temperature of the quantum annealing device, our results address the problem of the ``temperature scaling law for quantum annealers", which requires the temperature of quantum annealers to be reduced as problems of larger sizes are attempted to be solved.
\end{abstract}
\maketitle

\section{Introduction}

Quantum annealing (QA)~\cite{Apolloni:88,PhysRevB.39.11828,finnila_quantum_1994,kadowaki_quantum_1998,Brooke1999} is a predecessor of the quantum adiabatic algorithm (QAA) \cite{farhi_quantum_2000,Farhi:01} and adiabatic quantum optimization (AQO) \cite{Smelyanskiy:01,Reichardt:2004}, originally conceived as an algorithm that exploits simulated quantum (rather than thermal) fluctuations and tunneling to providing a quantum-inspired version of simulated annealing (SA) \cite{kirkpatrick_optimization_1983} for the solution of combinatorial optimization problems. Nowadays it is considered a special case of adiabatic quantum computation (AQC) \cite{Dam:2001fk}, a paradigm that is universal for quantum computation 
\cite{aharonov_adiabatic_2007,KempeGadget,MLM:06,OliveiraGadget,Aharonov:2009fi,Breuckmann:2014,Gosset:2014rp,Lloyd:2016}. For reviews see \cite{RevModPhys.80.1061,albash2016adiabatic}.

In AQC, the computation proceeds from an initial Hamiltonian whose ground state is easy to prepare, to a final Hamiltonian whose ground state encodes the solution to the computational problem. In the closed-system setting, the adiabatic theorem~\cite{Kato:50,Jansen:07} guarantees that the system will track the instantaneous ground state provided the Hamiltonian varies sufficiently slowly, and relates the required running time to the inverse of the minimum gap encountered during the computation. The only errors present in this setting are control errors and non-adiabatic transitions. The latter can be made arbitrarily small by a suitable choice of the path from the initial to the final Hamiltonian~\cite{lidar:102106,RPL:10,Wiebe:12,Ge:2015wo}. 

In the open-system setting, interactions with the environment produce additional errors in the form of dephasing between computational basis states and thermal excitations. Thermal excitation errors are suppressed due to the presence of finite gaps during the computation and corrected due to a certain amount of relaxation back to the ground state~\cite{childs_robustness_2001,PhysRevLett.95.250503,TAQC,Lloyd:2008zr,amin_decoherence_2009}. This picture is generally appropriate in the weak-coupling regime and when the temperature of the bath is small compared to the minimum gap \cite{Qiang:13,Albash:2015nx}. 

In practical implementations of AQC and QA, however, the system may be strongly coupled to the environment and the temperature of the thermal bath may be much larger than the small gaps expected when implementing computationally hard problems. It is not a surprise that in this scenario, AQC and QA require quantum error correction just like any other form of quantum information processing \cite{Lidar-Brun:book}. Moreover, it has been shown that for Hamiltonians that are sums of commuting two-body interactions it is not possible to even store quantum information reliably in the ground subspace (let alone compute), even at zero temperature \cite{Marvian:2014nr}. Unfortunately, despite the existence of various error suppression and correction techniques~\cite{jordan2006error,PhysRevLett.100.160506,PhysRevA.86.042333,Young:13,Sarovar:2013kx,Young:2013fk,Ganti:13,Mizel:2014sp,PAL:13,Young:2013fk,PAL:14,Vinci:2015jt,Mishra:2015,MNAL:15,Bookatz:2014uq,Marvian:2016kb,Marvian-Lidar:16,Marvian:2017aa}, it is not yet known how to achieve fault-tolerance in AQC and QA. Effective error suppression codes for AQC and QA typically require highly non-local Hamiltonians and very strong energy penalties \cite{Young:13,Sarovar:2013kx,Marvian:2016kb,Lloyd:2015fk,Marvian:2016kb,Marvian-Lidar:16,Marvian:2017aa} which are difficult to implement physically. Moreover, fault-tolerance assumes that it is possible to implement codes of arbitrary size (as measured in terms of the number of physical and logical qubits), in order to allow for scalable error correction of arbitrarily large computations. This is achieved in the circuit model, e.g., via concatenated codes~\cite{Aliferis:05}. While a comparable approach is still missing in AQC, nested quantum annealing correction (NQAC)~\cite{vinci2015nested} represents the first attempt to introduce scalable error correction for quantum annealing. NQAC is a scalable generalization of quantum annealing correction (QAC)~\cite{PAL:13,PAL:14,Vinci:2015jt,Mishra:2015,MNAL:15}, an error correcting techniques tailored to commercially available quantum annealing devices~\cite{Johnson:2010ys,Berkley:2010zr,Harris:2010kx,Dwave,Bunyk:2014hb}. 

Such devices operate in a noisy regime in which both thermal and analog control errors play  a crucial role in determining their performance~\cite{q108,SSSV,Albash:2014if,q-sig2,Crowley:2014qp,Martin-Mayor:2015dq,King:2015zr,Vinci:2014fk}. Indeed, both theoretical and empirical studies suggest current quantum annealing devices work in a \emph{quasi-static regime}~\cite{Amin:2015qf,Venuti:2015kq,marshall2017thermalization,chancellor2016maximum,Albash:2017ab}. In this regime, due to the strong interaction with the thermal bath, there is an initial phase of quasi-static evolution in which thermalization times are much shorter than the annealing time. Towards the end of the anneal, thermalization times grow and eventually become much longer than the annealing time. The system thus enters a  regime in which dynamics is frozen. The final outcomes of a quasi-static quantum annealing process thus provides a snapshot of the thermal (Gibbs) state of the system at the freezing point. The faster the freezing process, the closer to a thermal distribution at the freezing point the system will be at the end of the annealing process. A rough measure of performance for a quasi-static quantum annealer is thus given by an effective freezing temperature $T_{\rm eff}$, i.e., the temperature of the approximate thermal distribution at the freezing point.

Performing quantum annealing in a quasi-static regime may hide signs of potential quantum speedup, as discussed in Ref.~\cite{Amin:2015qf}. On the other hand, this provides an opportunity for using quantum annealers as thermal samplers that can be effectively used in machine-learning applications, e.g., in the training of Boltzmann machines~\cite{Adachi:2015qe,Amin:2016,benedetti2016quantum,korenkevych2016benchmarking}. On general grounds we expect the effective freezing temperature $T_{\rm eff}$ to grow with the problem size (earlier freezing or longer thermalization times for larger systems). The effects of intrinsic control errors typically also grows with problem size. Thus, no matter what the application, error correction remains crucial in order to ensure a scalable use of quantum annealers. 

A successful error correction scheme in quasi-static quantum annealing should effectively reduce the freezing temperature $T_{\rm eff}$:
\beq
T_{\rm eff} \rightarrow T_{\rm eff}/\mu\,, \quad \mu > 1\,.
\eeq
The temperature reduction can be equivalently interpreted as achieving an effective energy boost 
\beq
(h_i,J_{ij}) \rightarrow \mu (h_i,J_{ij})\,,
\eeq
where $h_i$ and $J_{ij}$ are the couplings defining the given problem Hamiltonian [see Eq.~\eqref{eq:HP}]. A scalable error correction scheme should allow for $\mu$ to scale with the code size. 

In Ref.~\cite{vinci2015nested} it was shown that NQAC indeed achieves error correction by introducing an energy boost $\mu$ and thus effectively reducing the effective temperature at which a D-Wave Two (DW2) quantum annealer operated. This immediately serves to address pessimistic ``temperature scaling law" conclusions of the type presented in Ref.~\cite{Albash:2017ab}, which do not account for error correction. Moreover, Ref.~\cite{vinci2015nested} showed that the energy boost obeys a power-law scaling:
\beq
\mu \sim \left(\frac{N_{\mathrm{phys}}}{N}\right)^\eta\, ,
\eeq
where $N_{\mathrm{phys}}$ is the number of physical qubits used to encode $N$ logical qubits.
These empirical results were theoretically interpreted in terms of a mean field analysis in Refs.~\cite{vinci2015nested,MNAL:15,MNVAL:PQAC}. 

Here we report our experiments with NQAC on a D-Wave 2000Q (DW2KQ) device. We briefly review NQAC in Sec.~\ref{sec:NQAC}. In Sec.~\ref{sec:OPT} we show that the power law scaling  found on a DW2 device continues to hold on a DW2KQ device, albeit with a reduced $\eta$ value. The DW2KQ device enabled us to use up to $728$ physical qubits for NQAC encoding, whereas NQAC on the DW2 was limited to $288$ physical qubits. We also show that NQAC outperforms classical repetition coding when NQAC is implemented in a regime in which the strength of the energy penalties is not a limiting factor. In Sec.~\ref{sec:SAMP} we show that NQAC can be implemented in conjunction with sampling applications. In particular, we show that NQAC can be used to reduce the sampling temperature of a quantum annealing device without significantly reducing the quality of the sampling. We conclude with a discussion of the implications of our results in Sec.~\ref{sec:DISC}.

\section{Nested Quantum Annealing Correction}
\label{sec:NQAC}

We start by briefly reviewing the nested quantum annealing correction (NQAC) construction~\cite{vinci2015nested}. A standard quantum annealing protocol is defined by the following transverse-field quantum annealing Hamiltonian:
\beq
H(t) = A(t) H_X + B(t)  H_{\mathrm{P}}\ , \qquad t\in[0,t_f] \ ,
\label{eq:adiabatic}
\eeq
in which the annealing schedules $A(t)$ and $B(t)$ (monotonically decreasing and increasing respectively) control the annealing schedule. The driver $H_X = -\sum_i \sigma_i^x$ 
serves to set up an initial uniform superposition (its ground state) and controls the tunneling rate, while the solution to an optimization problem of interest is encoded in the ground state of the Ising problem Hamiltonian $H_{\mathrm{P}}$:
\beq 
\label{eq:HP}
H_{\mathrm{P}} = \sum_{i \in \mc{V}} h_i \sigma^z_i + \sum_{(i,j) \in \mc{E}} J_{ij}\sigma^z_i\sigma^z_j\, .
\eeq
The  sums  above are performed over the vertices $\mc{V}$ and edges $\mc{E}$ of a graph $G = (\mc{V},\mc{E})$ that corresponds to the hardware (or connectivity) graph of a given quantum annealing device. The local fields $\{h_i\}$, couplings $\{J_{ij}\}$, and annealing time $t_f$ are programmed in order to represent the correct computational problem \cite{Johnson:2010ys,Berkley:2010zr,Harris:2010kx}. 

In quantum annealing correction, instead of trying to solve a computational problem by directly implementing the quantum annealing  Hamiltonian $H(t)$, one builds an ``encoded Hamiltonian" $\bar H(t)$~\cite{PAL:13,PAL:14,Vinci:2015jt,Mishra:2015,MNAL:15}:
\beq
\bar H(t) = A(t) H_X + B(t)  \bar H_{\mathrm{P}}\ , \qquad t\in[0,t_f] \ ,
\label{eq:encoded}
\eeq
defined over a set of physical qubits $N_{\rm phys }$ larger than the number of logical qubits $N$. We call $\bar H_{\mathrm{P}}$ the ``encoded problem Hamiltonian". The logical states of the logical Hamiltonian $H_{\mathrm{P}}$ are then recovered  through an appropriate decoding procedure of the states of $\bar H_{\mathrm{P}}$. Note that, due to practical limitations of current QA devices that prevent the programmability of the driver term $H_X$, only $H_{\mathrm{P}}$ is encoded in QAC.  

Nested quantum annealing correction was introduced in Ref.~\cite{vinci2015nested}%
\footnote{A $9$-bit instance of the scheme first appeared in Ref.~\cite{Young:2013fk} in the context of correcting uncorrelated random control errors on the problem Hamiltonian. This work showed that such control errors can be quadratically suppressed in the code distance using an NQAC-like scheme.}
with the goals of obtaining a QAC scheme that (1) can be implemented on arbitrary quadratic unconstrained optimization problems (QUBO), (2) is potentially scalable by allowing a variable code-size and (3) could be implemented on a generic quantum annealing device. These three goals were achieved by means of a ``nesting" procedure that we now outline. 

The most general QUBO has arbitrary pairwise interactions between a set of $N$ logical variables. We thus consider the encoding of a problem Hamiltonian $H_{\mathrm{P}}$ defined on a complete graph $K_N$. NQAC involves two maps between three types of qubits: (i) each logical qubit (representing a logical variable) is mapped to a set of code qubits, whereupon $H_{\mathrm{P}} \mapsto \tilde H_{\mathrm{P}}$, (ii) each code qubit is mapped to a set of physical qubits, whereupon $\tilde H_{\mathrm{P}} \mapsto \bar H_{\mathrm{P}}$, the encoded problem Hamiltonian in Eq.~\eqref{eq:encoded}. The role of the code qubits is primarily to provide protection against thermal and control errors. The code qubits typically comprise a repetition code, though more general stabilizer codes are certainly possible in principle~\cite{Young:2013fk}. The map from code qubits to physical qubits is required for embedding purposes, since the code qubits of a given logical qubit are fully connected in NQAC, and the hardware graph $G$ is typically not fully connected. The repetition code can be decoded at the end of each annealing run, which means that excited states can be used profitably to recover the sought-after ground state of $H_{\mathrm{P}}$. We now describe each of these components of NQAC in turn, in more detail

\subsection{From logical qubits to code qubits}
The first step of the NQAC scheme consists in transforming the logical problem $H_{\mathrm{P}}$ into a ``nested"  Hamiltonian $\tilde H_{\mathrm{P}}$ that is defined on a larger $K_{C\times N}$.  The nesting level $C$ controls the amount of hardware resources (qubits, couplers, etc.) used in the transformation and enables the  scalability of the error correction method. Each logical qubit $i$ ($i = 1,\dots,N$) in  $\tilde H_{\mathrm{P}}$  is represented by a $C$-tuple of code qubits $(i,c)$, with $c = 1,\dots,C$. This $C$-tuple occupies the vertices of a $K_C$ with equally weighted  ferromagnetic edges, which also serves as a distance-$C$ repetition code. Code qubits $c$ and $c'$ belonging to different logical qubits $i$ and $j$ are coupled with strength $J_{ij}$. Thus, the ``nested" couplers $\tilde J_{(i,c),(j,c')}$ and local fields $\tilde h_{(i,c)}$ are defined as follows: 
\bes
\label{eq:nesting}
\begin{align}
\tilde J_{(i,c),(j,c')} &= J_{ij}\,, \quad  \forall c,c', i\neq j\ ,   \\
\tilde h_{(i,c)} &= C h_{i}\,, \quad  \forall c, i\ , \label{eqt:h} \\
\tilde J_{(i,c),(i,c')} &= -\gamma \,, \quad  \forall c\neq c' \ .
\end{align}
\ees
Note that each logical coupling  $J_{ij}$ has $C^2$ copies $\tilde J_{(i,c),(j,c')}$ while the local fields $h_{i}$ have $C$ copies $\tilde h_{(i,c)}$. For each logical qubit $i$, there are $C(C-1)/2$ ferromagnetic couplings $\tilde J_{(i,c),(i,c')}$ of strength $\gamma>0$ representing energy penalties that facilitate the alignment of the $C$ code qubits that represent one logical qubit.

\subsection{From code qubits to physical qubits}
The nested Hamiltonian $\tilde H_{\mathrm{P}}$ constructed over the code qubits in the previous step must be implemented in given QA hardware. The required transformation from code qubits to physical qubits, i.e., from $\tilde H_{\mathrm{P}}$ to $\bar H_{\mathrm{P}}$ can be accomplished using, e.g., minor embedding (ME)~\cite{Kaminsky-Lloyd,Choi2,klymko_adiabatic_2012,Cai:2014nx,Boothby2015a}, or the Lechner-Hauke-Zoller (LHZ) scheme~\cite{Lechner:2015,albash2016simulated,Pastawski:2015}. We focus here on the ME  scheme primarily since it is the relevant one for D-Wave quantum annealers. 

The ME  step replaces each code qubit in $\tilde H_{\mathrm{P}}$ by a ferromagnetically coupled chain of physical qubits, such that all couplings in $\tilde H_{\mathrm{P}}$ are represented by inter-chain couplings.  The intra-chain coupling represents additional energy penalties that force the chain of physical qubits to behave as a single code qubit. 
The minor embedding of a $K_{C\times N}$ graph requires that each code qubit $(i,c)$ is represented by a physical chain of length $L = \lceil C N/4 \rceil+1$ on the Chimera graph \cite{Choi2}. The number of physical qubits necessary to implement the $C$-th level of nested encoding (i.e., a distance-$C$ code) of a problem with $N$ logical variables is thus  
\beq
N^{\mathrm{phys}}_{C} =  CNL \sim C^2N^2/4\ ,
\eeq 
i.e., it scales quadratically with both $C$ and $N$.

\subsection{Decoding}
Finally, a decoding procedure must be employed to recover the logical state from the readout of the physical qubits. This is a two-step process since we must first decode the length-$L$ chain of physical qubits representing each code qubit $(i,c)$, and then decode the $C$ code qubits to yield the state of the $i$-th logical qubit. For simplicity we only consider majority vote decoding for both steps, although other decoding strategies are possible and have been explored~\cite{Vinci:2015jt}.

\subsection{Boosting effect}
As mentioned above, error correction for quantum annealing in the quasi-static scenario should achieve a scalable effective temperature reduction. In the NQAC construction, each logical coupling is represented by $C^2$ physical copies and could ideally provide a maximal energy boost that grows quadratically with the nesting level $C$: $\mu \sim C^\eta$ with $\eta=2$. A mean-field analysis has confirmed this by showing that the free energy $\mathcal F$ after nesting is indeed equal to the free energy without encoding provided the logical couplings are boosted by a factor of $C^2$~\cite{MNAL:15,vinci2015nested,MNVAL:PQAC}:
\beq
\mathcal F_C(\beta,J,\lambda, \Gamma) = \mathcal F_1(\beta,C^2J,C^2\lambda,C\Gamma)\,.
\eeq
Note, however, that the transverse field is only sub-optimally boosted by a factor of $C$: $\Gamma\mapsto C\Gamma$. This is due to the aforementioned limitation of not being able to  encode the driver Hamiltonian. Despite this limitation, NQAC was shown to provide a scalable energy boost in experiments performed with a DW2 quantum annealer, in which small logical problems were nested until the physically encoded Hamiltonian used up to $288$ physical qubits. The empirical findings confirmed a power law scaling of the energy boost with nesting level, but with a suboptimal scaling factor $\eta <2$. This suboptimal scaling is a consequence of the combined detrimental effects of minor embedding, control errors, and the role played in the dynamics by the unencoded transverse field~\cite{vinci2015nested}.

\begin{figure*}[ht]
\begin{center}
\subfigure[\,]{\includegraphics[width=0.33\textwidth]{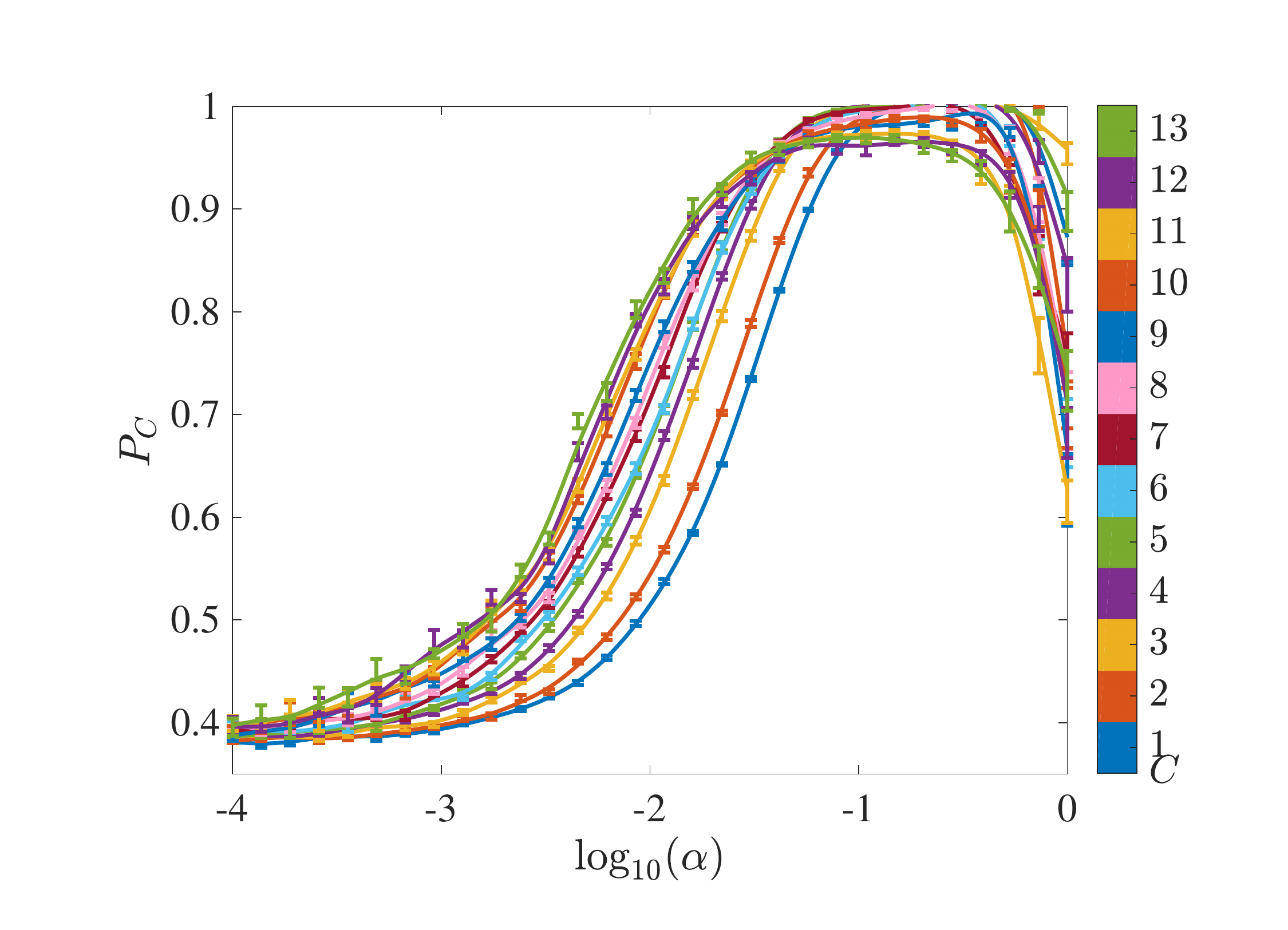}\label{fig:1a}}
\subfigure[\,]{\includegraphics[width=0.33\textwidth]{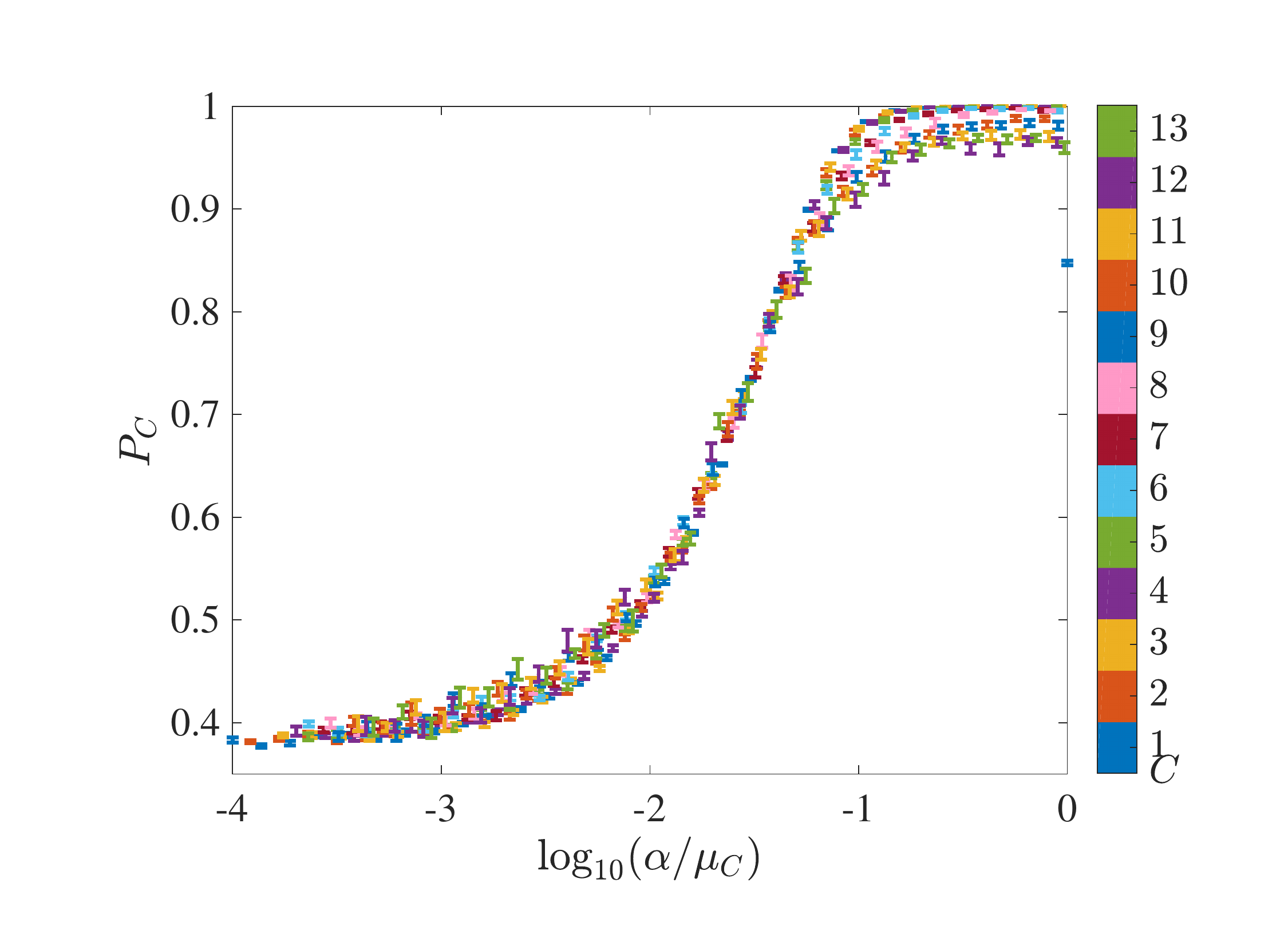}\label{fig:1b}}
\subfigure[\,]{\includegraphics[width=0.33\textwidth]{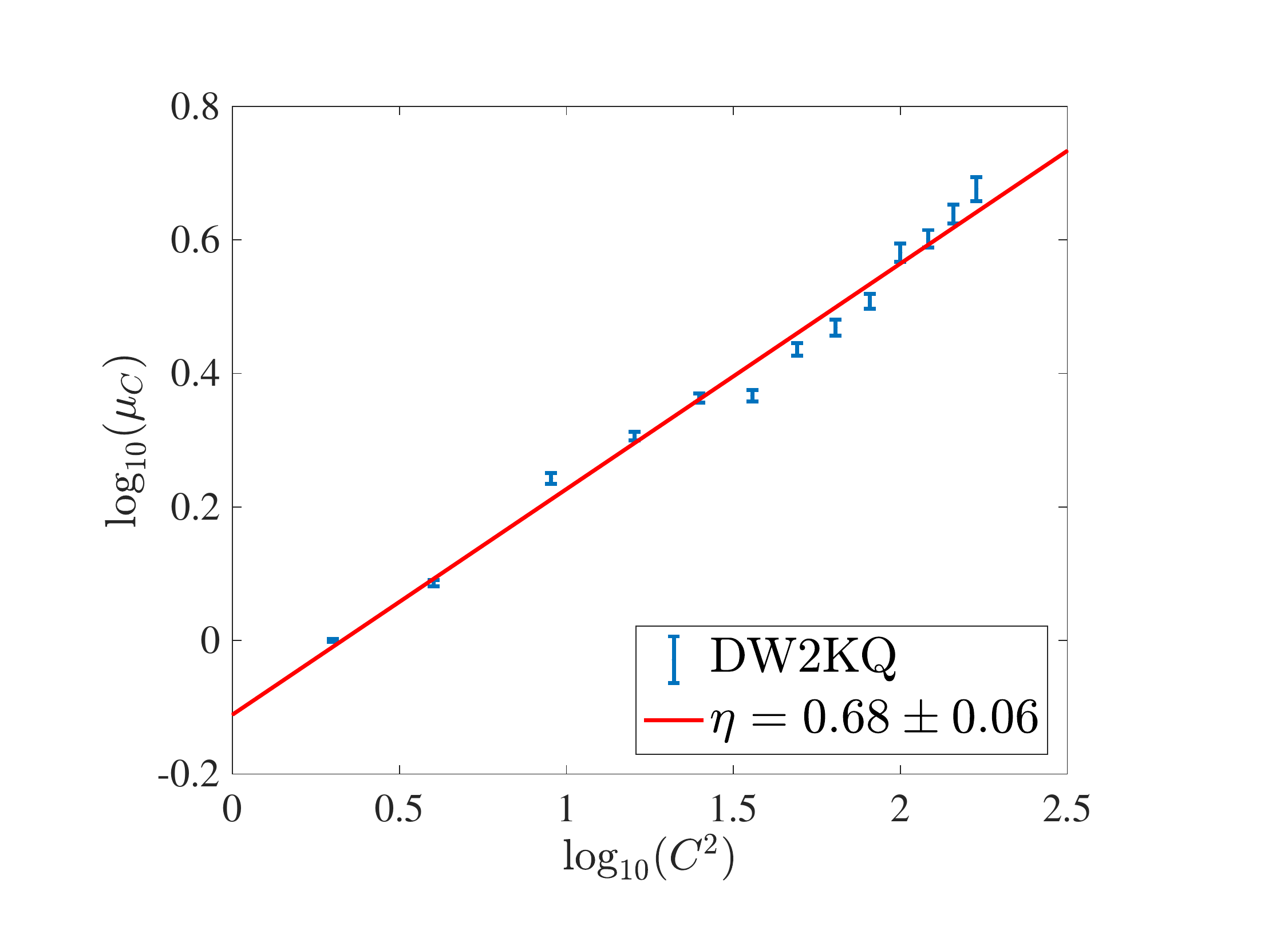}\label{fig:1c}}
\caption{Empirical results for the antiferromagnetic $K_4$, after encoding, followed by ME and decoding. (a) DW2KQ success probabilities $P_C(\alpha)$ for 
$13$ nesting levels $C$. Increasing $C$ generally increases $P_C(\alpha)$ at fixed $\alpha$. (b) Rescaled $P_C(\alpha)\mapsto P_C(\alpha/\mu_C)$ data, exhibiting data-collapse. Here $\mu_C$ is chosen for each $C$ such that the quality of the data-collapse is maximized. Using the values of $\mu_C$ found from data-collapse, (c) shows the scaling of the energy boost $\mu_C$ \textit{vs} the amount of physical resources used ($N_{\rm phys} \sim C^2$). Empirical data are consistent with a power-law scaling $\mu_C \sim C^\eta$, where $\eta = 0.68\pm0.6$. In all panels $N_{\mathrm{phys}}\in [8,728]$. The error bars in the left and center figures correspond to the median, 25-th and 75-th percentiles of $P_C(\alpha)$ computed for each embedding.  See Appendix~\ref{app:B} for more details on the determination of $\mu_C$ and the associated error bars. }
\label{fig:1}
\end{center}
\end{figure*}

\section{NQAC for optimization}
\label{sec:OPT}

In this and the next section we discuss NQAC in experiments with up to $728$ physical qubits performed on a DW2KQ device, up from $288$ reported in Ref.~\cite{vinci2015nested}, which used a DW2 device. The results of this section confirm and extend those obtained in Ref.~\cite{vinci2015nested}, while the results of the next section address NQAC in the new context of sampling. Experimental background details are given in Appendix~\ref{app:A}.

\subsection{Scalable energy boost}

We closely follow the methodology of Ref.~\cite{vinci2015nested} to show that NQAC provides an energy boost that scales as a power law with the number of physical qubits used in the encoding. 

We start by introducing an overall energy scale $\alpha$~\cite{q-sig2} for the problem Hamiltonian: $H_{\mathrm{P}} \mapsto \alpha H_{\mathrm{P}}$, with $0<\alpha\leq 1$. The dimensionless parameter $\alpha$ controls the level of thermal and dynamically induced errors, as well as the relevance of analog control errors, and thus controls the hardness of an optimization problem when implemented on a quantum annealing device. The introduction of the scale $\alpha$ also allows us to extract the energy boost $\mu_C$ provided by an NQAC encoding with $C$ levels of nesting. The energy boost $\mu_C$ is defined as the rescaling factor required so that
\beq
P_C(\alpha) \approx P_1(\mu_C \alpha) \,,
\label{eq:mu_C-def}
\eeq
where $P_C$ is the success probability of QA when a given optimization problem is implemented with $C$ levels of nesting, and $C=1$ corresponds to the unencoded case.  
Equation~\eqref{eq:mu_C-def} states that the performance enhancement obtained at nesting level $C$ can be interpreted as increasing the energy scale $\alpha$ by a factor of $\mu_C$ relative to the unencoded case.

We are interested in the scaling of  $\mu_C$ for large $C$. We thus consider the implementation  of a small logical problem that can be nested several times. On the DW2KQ device, a completely antiferromagnetic ($h_i=0$, $\forall\, i$) Ising problem over $K_4$ ($J_{ij} = 1$ $\forall\, i,j$) can be nested up to $C=13$ with balanced minor embeddings.%
\footnote{In a balanced embedding all chains have same length. We limited our experiments to balanced embedding since we found that the use of more general, unbalanced, embeddings reduces the performance enhancement provided by NQAC.} 
The results are shown in Fig.~\ref{fig:1}. Figure~\ref{fig:1a} shows the success probability $P_C(\alpha)$ as a function of the nesting level $C$ and of the energy scale $\alpha$.  The strength of the energy penalty $\gamma$ was optimized to obtain the best performance for the NQAC encoding among the 9 values $\gamma \in \{0.1,0.2,\dots,0.9,1\} $. For all experiments we  used $25$ different, randomly generated balanced embeddings using the algorithm introduced in Ref.~\cite{Boothby2015a}.  For each embedding we performed $1000$ annealing runs. As expected, $P_C(\alpha)$ drops from $P_C(1) = 1$ (solution always found) to $P_C(0) = 6/16$ (random sampling of $6$ ground states, where $4$ out of the $6$ couplings are satisfied, out of a total of $16$ states). 

In Fig.~\ref{fig:1a} we see two regions in which NQAC has a  different behavior: small and large $\alpha$. In the small $\alpha$ region the success probability grows monotonically with $\alpha$ and $C$. In the large $\alpha$ region the success probability starts dropping for large $C$. This drop is due to the fact that the largest implementable value for the energy penalty $\gamma$ is suboptimal, i.e., too small. For this reason, we refer to the large $\alpha$ region as ``penalty-limited". 
The monotonic improvement of the success probability in the small $\alpha$ region is consistent with a scalable increase of $\mu_C$. We thus refer to the small $\alpha$ region as the ``scaling" region. 

Figure~\ref{fig:1b}  shows that the data from the left panel can be collapsed using the scaling ansatz given in Eq.~\eqref{eq:mu_C-def}, and hence that the effects of nesting can be interpreted, at least in the scaling region, as providing an energy boost $\mu_C$. Figure~\ref{fig:1c} shows the energy boost $\mu_C$ as a function of $C$, as determined via the data collapse shown in the middle panel. We see that the energy boost $\mu_C$ grows monotonically with $C$ following the power law scaling  $\mu_C\sim C^\eta$ with $\eta$ determined empirically to be $\eta \sim 0.68$. 

We have previously observed that a nested graph $K_{C\times N}$ contains $C^2$ equivalent copies of the same logical coupling $J_{ij}$, thus intuitively providing a maximal energy boost in which $\mu_C\sim C^2$. We empirically find $\eta \approx 0.68 < 2$. This suboptimal scaling was explained in Ref.~\cite{vinci2015nested} as being attributable mainly to the overhead cost of minor embedding and the detrimental effect of control errors. The scaling parameter $\eta$ was found to be $\eta \approx 1.1 < 2$ on a DW2 device. We expect to observe different scaling due to the various differences between the DW2 and DW2KQ devices: smaller operating temperature for the DW2KQ, smaller analog control error strength for the DW2KQ, different embedding used for NQAC due to uncalibrated qubits (fewer in DW2KQ but more uniformly distributed over the hardware graph). 

\begin{figure}[ht]
\begin{center}
\subfigure[\, $K_{16}$ ] {\includegraphics[width=0.25\textwidth]{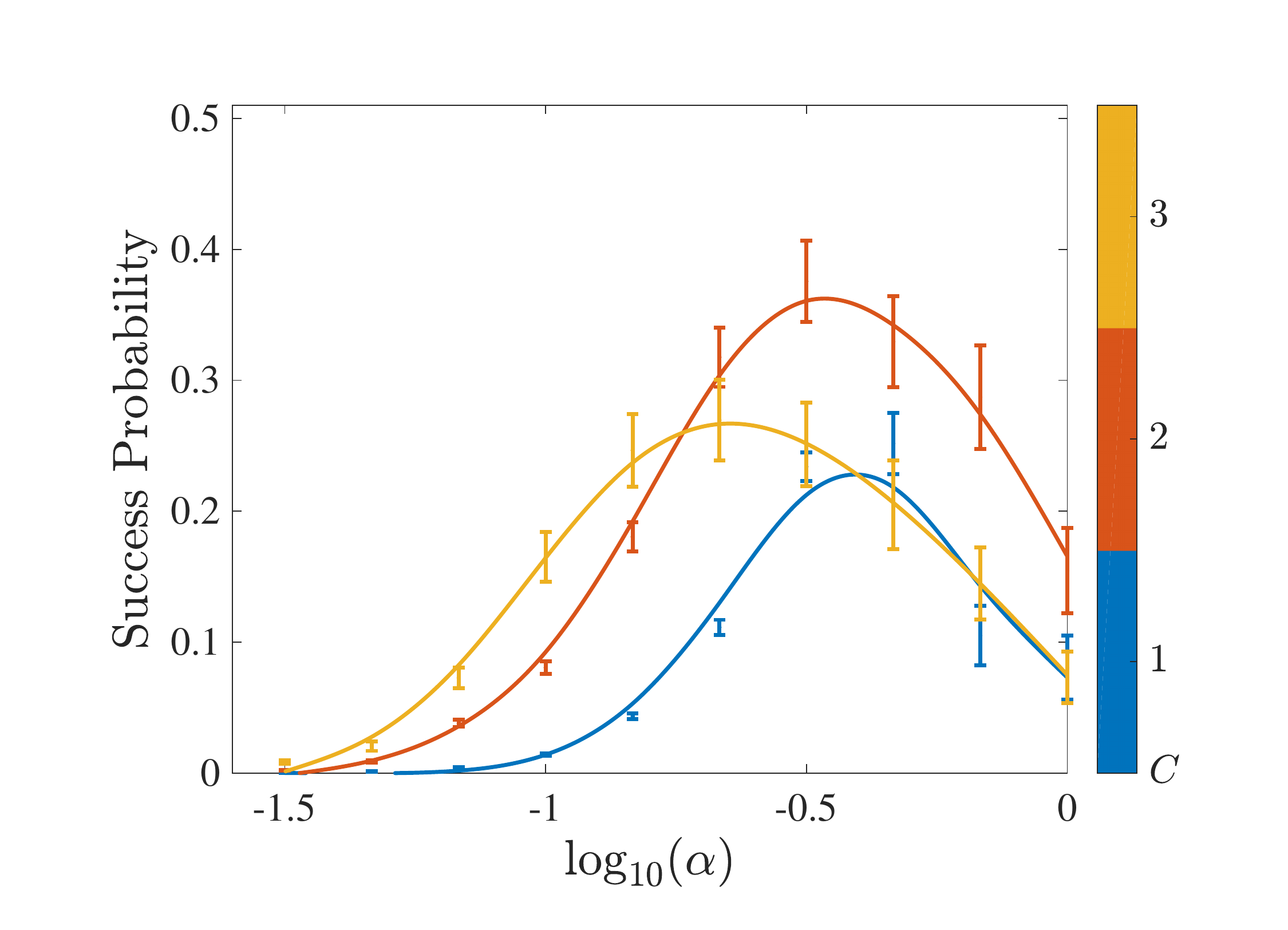}\label{fig:2a}}\hspace{-.6cm}
\subfigure[\,$K_{16}$]{\includegraphics[width=0.25\textwidth]{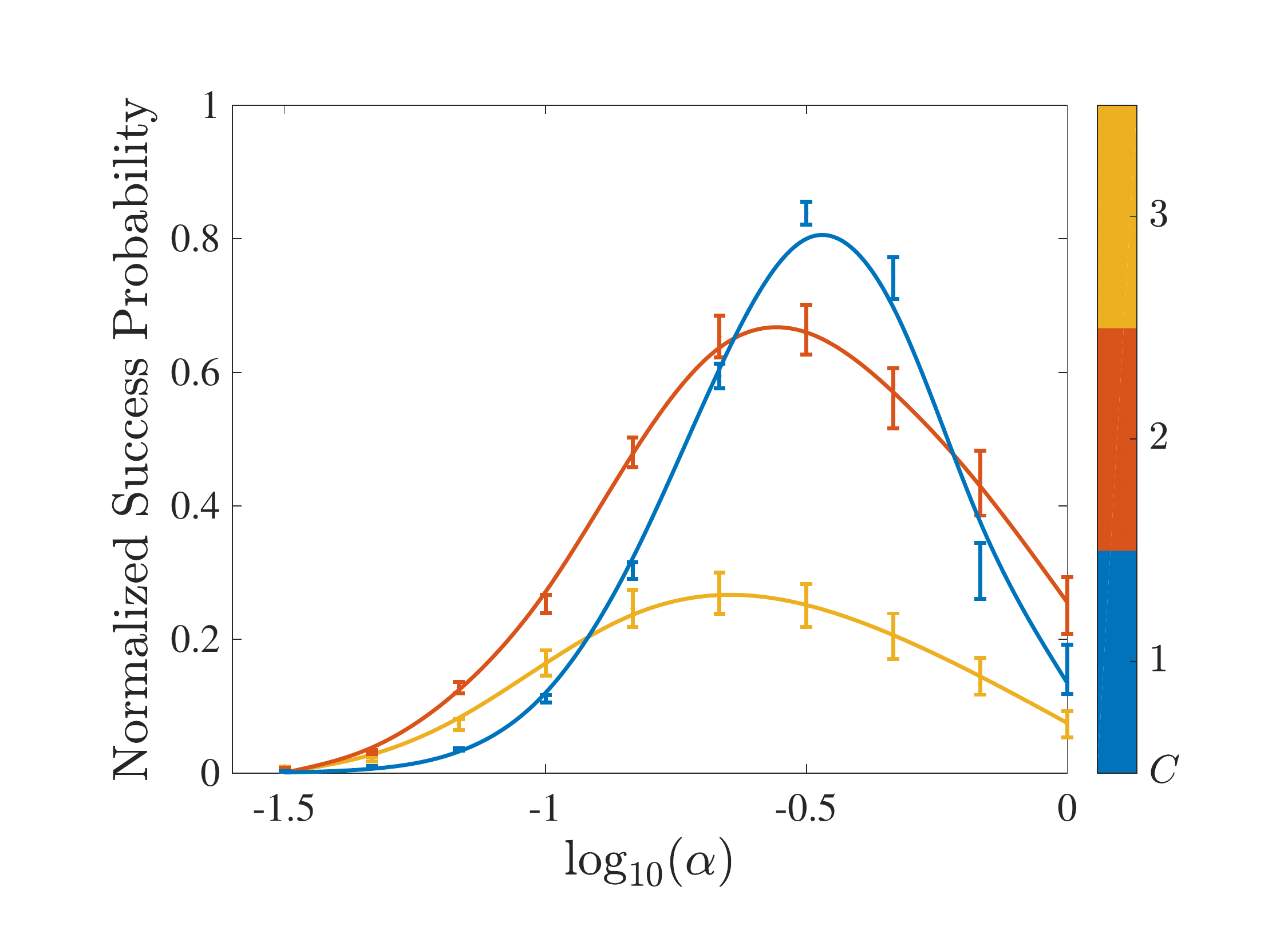}\label{fig:2b}}\\
\subfigure[\,$K_{16}$]{\includegraphics[width=0.25\textwidth]{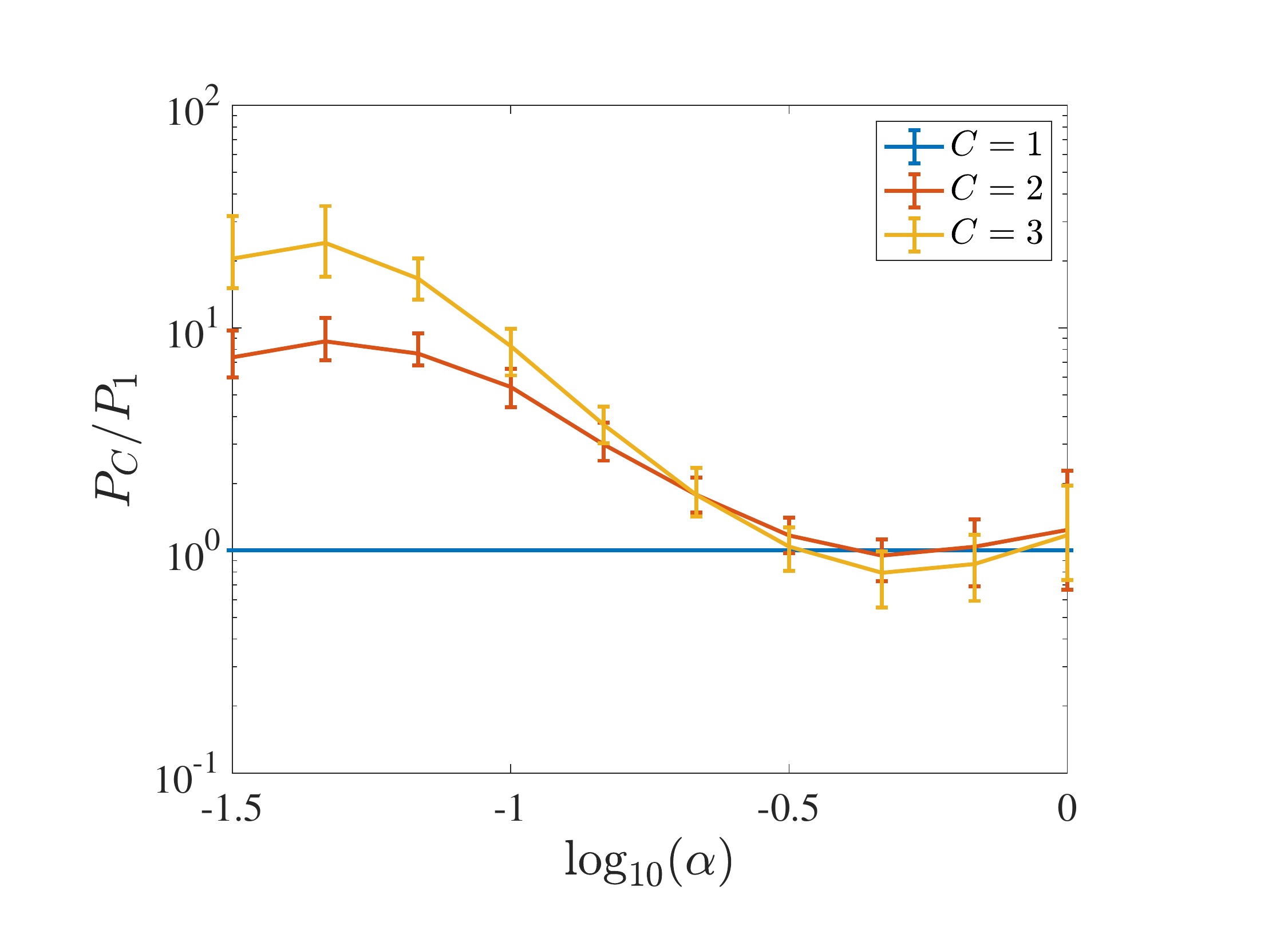}\label{fig:2c}}\hspace{-.6cm}
\subfigure[\,$K_{16}$]{\includegraphics[width=0.25\textwidth]{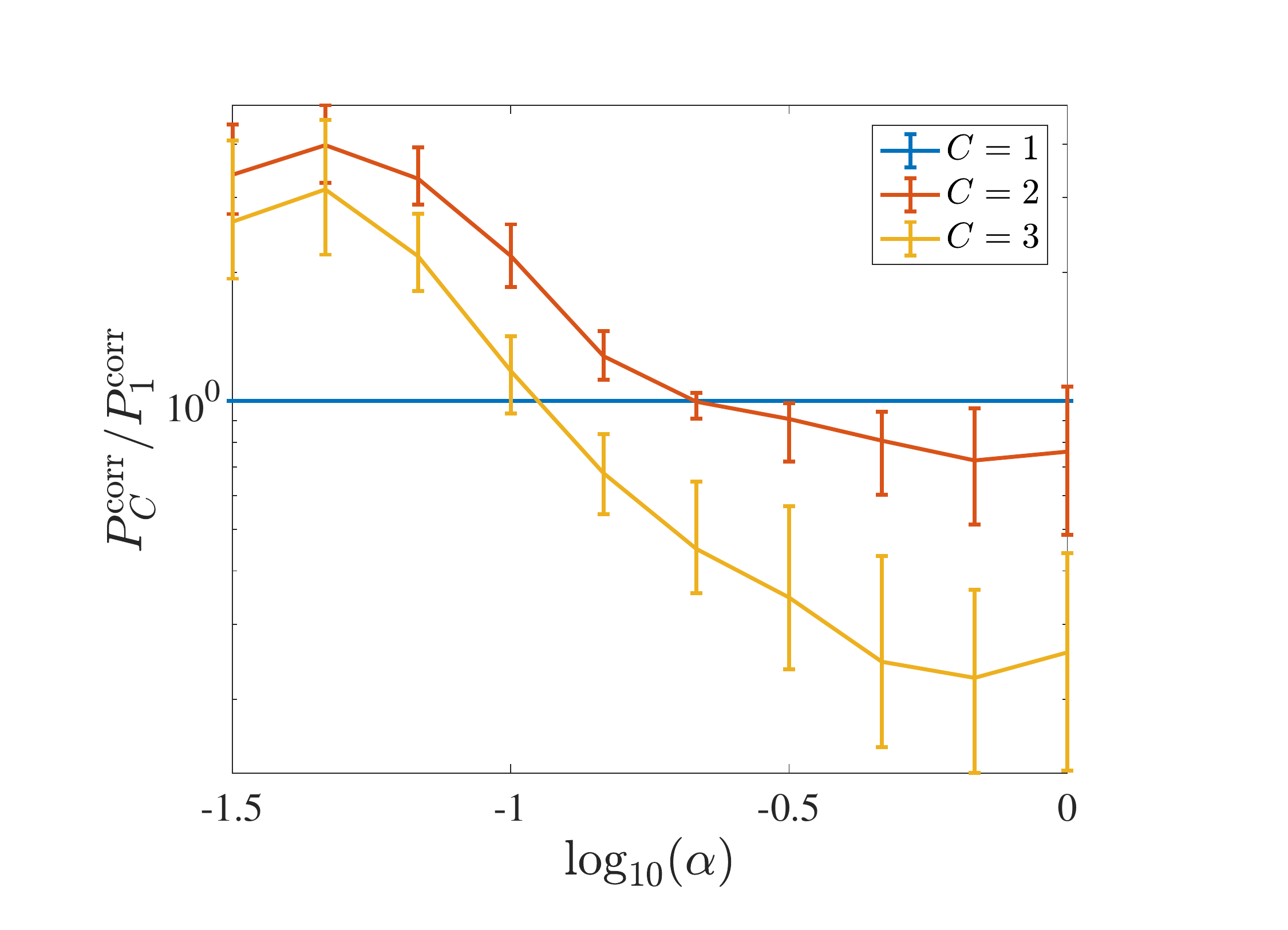}\label{fig:2d}}\\
\subfigure[\,$K_{24}$]{\includegraphics[width=0.25\textwidth]{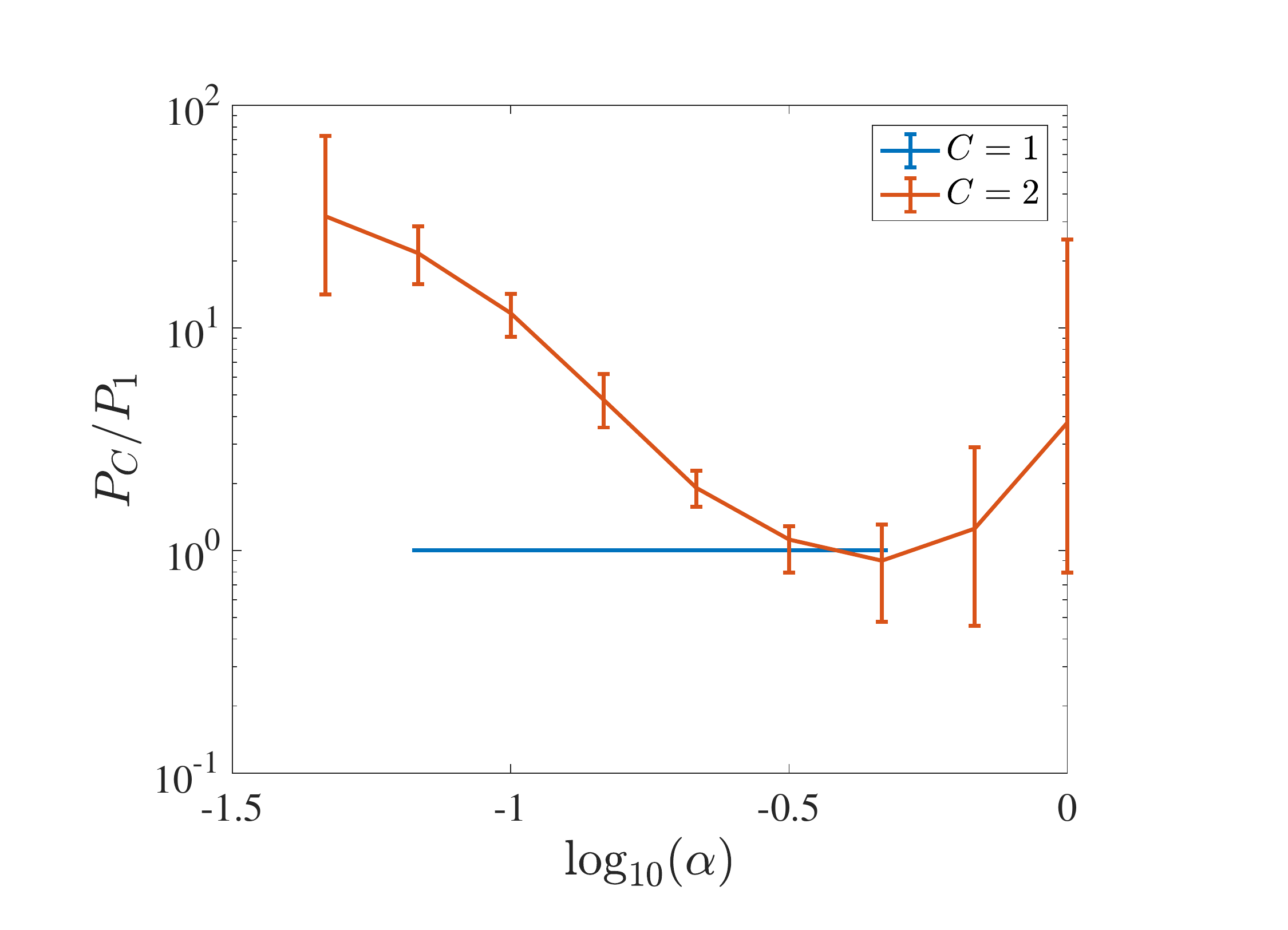}\label{fig:2e}}\hspace{-.6cm}
\subfigure[\,$K_{24}$]{\includegraphics[width=0.25\textwidth]{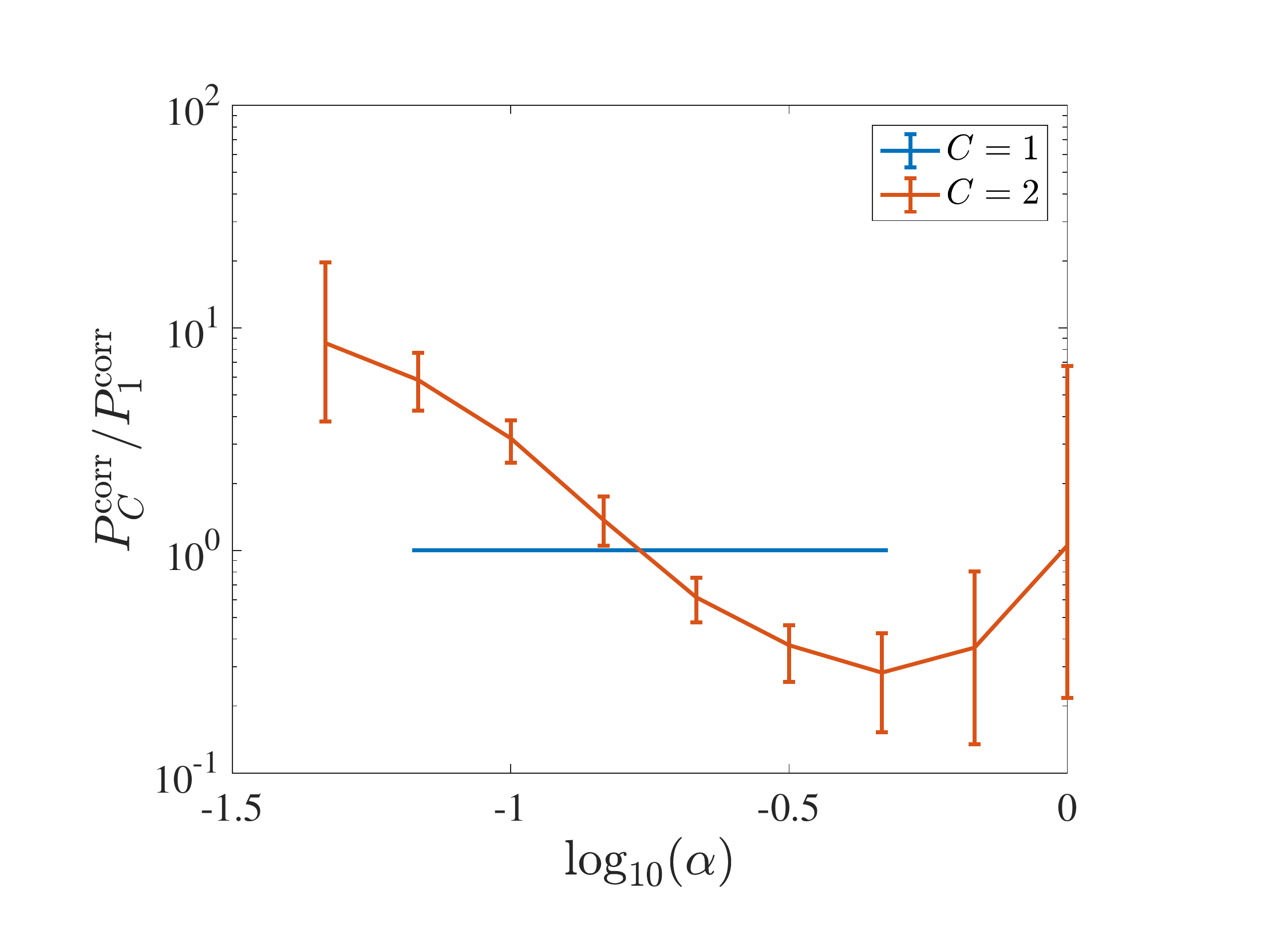}\label{fig:2f}}
\caption{(a) Success probabilities for a random $K_{16}$ instance with up to $C=3$ nesting levels. Note the monotonic and non-monotonic improvements of $P_C$ in the scaling (small $\alpha$) and penalty-limited (large $\alpha$) regions. (b) Same as (a), but with success probabilities accounting for classical repetition via Eq.~\eqref{eq:P_C-para}. Classical repetition outperforms NQAC in the penalty-limited region while there is an optimal nesting level ($C=2$) for NQAC in the scaling region. Panels (c) and (d) display the normalized success probabilities without (c) and with (d) classical repetition for the $K_{16}$ ensemble. Panels (e) and (f) display the same for the $K_{24}$ ensemble. The behavior of the example instance in (a) and (b) is representative for the class of random instances it belongs to.} 
\label{fig:2}
\end{center}
\end{figure}

\subsection{Classical repetition}

The existence of two separate regions (scaling and penalty-limited) becomes more evident when we consider the encoding of larger logical problems. In Fig.~\ref{fig:2a} we show the empirical results for an example instance (with couplings randomly generated from the set $J_{ij} \in \{\pm 0.1,\pm 0.2,\dots,\pm 0.9,\pm 1\}$) defined on a $K_{16}$ with up to $C=3$ nesting levels. In the scaling region the success probability is monotonically increasing. In the  penalty-limited region the success probability peaks at a given value where the optimal penalty strength is close to the largest allowed value and then decreases again at larger values of $\alpha$. Moreover, the success probability is typically not a monotonically increasing function of the nesting level $C$. 

The presence of a penalty-limited region prevents NQAC from providing a scalable energy boost for a wider range of energy scales $\alpha$. This effects is particularly important when comparing NQAC to classical repetition.  For NQAC to be considered as a practically useful scheme for error correction in quantum annealing, it must be more effective than a classical repetition scheme that uses the same amount of physical resources as the NQAC encoding. In particular, let us denote by ${C_{\max}}$ the largest nesting level $C$ that can be implemented on the QA device.   Then $M_C = \lfloor N^{\mathrm{phys}}_{C_{\max}}/N^{\mathrm{phys}}_{C}\rfloor$ is the number of copies that can be implemented in parallel for a given nesting level $C$. The ``corrected" success probability of an NQAC encoded optimization problem is thus the probability of succeeding at least once after a run with $M_C$ statistically independent, parallel copies: 
\beq
P_C^{\rm corr}(\alpha) = 1-[1-P_C(\alpha)]^{M_C}\ . 
\label{eq:P_C-para}
\eeq
Figure~\ref{fig:2b} shows the parallelized success probability for the example instance of Fig.~\ref{fig:2a}. We see that nesting does not help in the penalty-limited region (the unencoded $C=1$ case gives the best results). However, nesting does help in the scaling region, where the strength of the energy penalties is not a bottleneck. In the case of the example shown we find the existence of an optimal nesting level $C=2$ in the scaling region.

We now show that the empirical results discussed regarding the example instance shown in Figs.~\ref{fig:2a} and~\ref{fig:2b} are representative of the general behavior of NQAC. We have confirmed this by studying four ensembles of $100$ randomly generated instances on $16$ and $24$ logical variables ($K_{16}$ and  $K_{24})$. Couplings are randomly generated from the set $J_{ij} \in \{\pm0.1,\pm0.2,\dots,\pm0.9,\pm1\}$. As in the previous section we used $25$  randomly generated balanced embeddings and for each embedding we performed $1000$ annealing runs. To keep the running time on the DW2KQ manageable, we did not optimize the energy penalty $\gamma$, which we fixed to its maximum value $\gamma = 1$. Figures~\ref{fig:2c}-\ref{fig:2f} show the experimental results. The figures show the median (error bars represent the $25$th and $75$th percentiles) ratios ($P_C/P_1$) and ($P^{\rm corr}_C/P^{\rm corr}_1$). Nesting is preferable when the $C>1$ curves are above the  horizontal breakeven line. Figures~\ref{fig:2c} and~\ref{fig:2e} show that nesting generically provides a success probability improvement that is monotonically increasing with $C$ in the scaling region (small $\alpha$), while it does not necessarily improve performance in the penalty-limited region. Figures~\ref{fig:2d} and~\ref{fig:2f} show that nesting typically outperforms the unencoded ($C=1$) case in the scaling regime, while nesting does not help in the penalty-limited region. After correcting for classical repetition, the success probability improvement is not necessarily monotonic in the scaling region. For example Fig.~\ref{fig:2d} shows that $C=2$ is optimal for $K_{16}$, in the sense that $P^{\rm corr}_2 > P^{\rm corr}_1,P^{\rm corr}_3$ .

\section{NQAC for sampling}
\label{sec:SAMP}

In this section we explore NQAC in the context of quantum annealing used for sampling applications. The limited hardware connectivity of quantum annealing devices and the finite strength and precision of their couplings are major limitations not only for optimization (previous section), but also for sampling applications. We are particularly interested in machine-learning applications in which quantum annealers are used as samplers for training Boltzmann machines~\cite{Adachi:2015qe,Amin:2016,benedetti2016quantum,korenkevych2016benchmarking}. 

\subsection{Training Boltzmann machines}

Boltzmann machines are generative probabilistic models that can be used for both supervised and non-supervised machine-learning applications. Boltzmann machines can be used as a building block for deep networks thus playing a role in the booming fields of artificial intelligence and deep-learning. A Boltzmann machine associates a given data point $z\equiv \{z_i\}$ (here represented as a string $z_i = \pm 1, i = 1,...,N$) to an Ising ``energy" function $E(z)$
\beq
E(z) = \sum_{i \in  \mathcal V} b_i z_i+  \sum_{(i,j) \in \mathcal E}  w_{ij}z_iz_j\,,
\label{eq:BE}
\eeq
and a corresponding probability distribution $P(z)$: 
\beq
P(z) = e^{-E(z)}/Z,\quad Z = \sum_z e^{-E(z)}\ .
\label{eq:BP}
\eeq
This is a Boltzmann distribution with unit temperature [the connection to the physical temperature is discussed below --- see Eq.~\eqref{eq:bi-wij}].
The energy function in Eq.~\eqref{eq:BE} corresponds to an Ising model defined on a graph $G = (\mathcal V ,\mathcal E )$. A Boltzmann machine is thus also a ``graphical model". Training a Boltzmann machine consists of finding the values of the weights $b_i$ and $w_{ij}$ such that the probability distribution $P(z)$ generated by the model will extract and generalize as many ``features" as possible from the data set. The training of a Boltzmann machine is achieved by minimizing the negative log-likelihood of the  data set with respect to the probability distribution $P(z)$. This results in  iteratively adjusting the weights of the model according to the following update rules:
\bea
\delta b_i & \sim & \langle  z_i \rangle_{\rm data} -  \langle  z_i \rangle_{T}\ \nonumber \\
\delta w_{ij} & \sim & \langle  z_i z_j \rangle_{\rm data} -  \langle  z_iz_j \rangle_{T}\,,
\label{eq:grads}
\eea
where the first terms are averages over the data set and are usually called ``positive phases" while the second terms are thermal averages [i.e.,  $ \langle  X \rangle_{T}\equiv \sum_i x_i P(x_i)$ where $x_i$ are the values taken by the random variable $X$], and are usually called ``negative phases". Computing the negative phases is known to be computationally hard with classical algorithms, essentially since it requires sampling from the Gibbs state of a spin glass, and is usually replaced with the contrastive divergence approximation \cite{Hinton:2006aa,Bengio:2009aa}. Instead, one may try to compute the negative phases using physical quantum annealing devices as Boltzmann samplers ~\cite{Adachi:2015qe,Amin:2016,benedetti2016quantum,korenkevych2016benchmarking}. The hope is that this can be done faster than a classical computation of the associated partition function, which would be an example of a quantum advantage in quantum machine learning \cite{Biamonte:2016aa}.

\subsection{Relating the Boltzmann machine parameters to the quantum annealer parameters}

As we discussed previously, performing quantum annealing in a quasi-static regime allows one to approximately sample from the instantaneous thermal state at the freezing point \cite{Amin:2015qf}. We shall assume here that the freezing point happens late enough in the anneal such that the system freezes when the problem Hamiltonian $H_{\rm P}$ dominates. Performing quantum annealing with a device that realizes Eqs.~\eqref{eq:adiabatic} and~\eqref{eq:HP} allows to thermally sample from the Boltzmann distribution of the model given by Eqs.~\eqref{eq:BE} and~\eqref{eq:BP}. 

Let $\beta_{\rm phys}$ denote the inverse physical temperature of the quantum annealing device, $B(t)$ the profile function controlling the energy scale of the problem Hamiltonian $H_{\rm P}$ [as in Eq.~\eqref{eq:adiabatic}], and $t^*<t_f$ the freezing time. Because of freezing, the quantum annealer (approximately) samples from a Gibbs distribution at an effective temperature $T_{\rm eff} = 1/\beta_{\rm eff}$ given by:
\beq
\beta_{\rm eff} = \beta_{\rm phys} B(t^*)/B(t_f)\,.
\eeq
The existence of an effective sampling temperature allows us to establish the connection between the physical parameters of a quantum annealing device as defined in Eq.~\eqref{eq:HP} and the weights of a Boltzmann machine as defined in Eq.~\eqref{eq:BE}:
\beq
b_i= \beta_{\rm eff} h_i,\quad  w_{ij}= \beta_{\rm eff} J_{ij}\,.
\label{eq:bi-wij}
\eeq

\subsection{Methodology}
\label{sec:method}

Our goal in this section is to study whether NQAC can be used to improve the performance of quantum annealers in the training of fully connected (as opposed to restricted \cite{Bengio:2009aa}) Boltzmann machines, i.e., when the graphical model underlying the model of Eq.~\eqref{eq:BE} is defined on a complete graph $K_N$. Due to the physical implementation of quantum annealing on the Chimera graph, this means that the Boltzmann machine is necessarily defined on the logical problem Hamiltonian $H_P$ [Eq.~\eqref{eq:adiabatic}].
In order for NQAC to be successful for sampling it should both allow for reliably sampling from logical thermal distributions (defined over the logical problem Hamiltonian), and result in a reduction of the effective temperature $T_{\rm eff}=1/\beta_{\rm eff}$ [or, equivalently, an effective boost of the annealing couplers $(h_i , J_{ij})$]. Of course, this is accomplished by sampling from the corresponding minor embedded implementations defined via the encoded Hamiltonian $\bar H_P$ [Eq.~\eqref{eq:encoded}].

\begin{figure*}[ht]
\begin{center}
\subfigure[\,$K_{8}$: $\beta_{C,\rm eff}$, unscaled]{\includegraphics[width=0.33\textwidth]{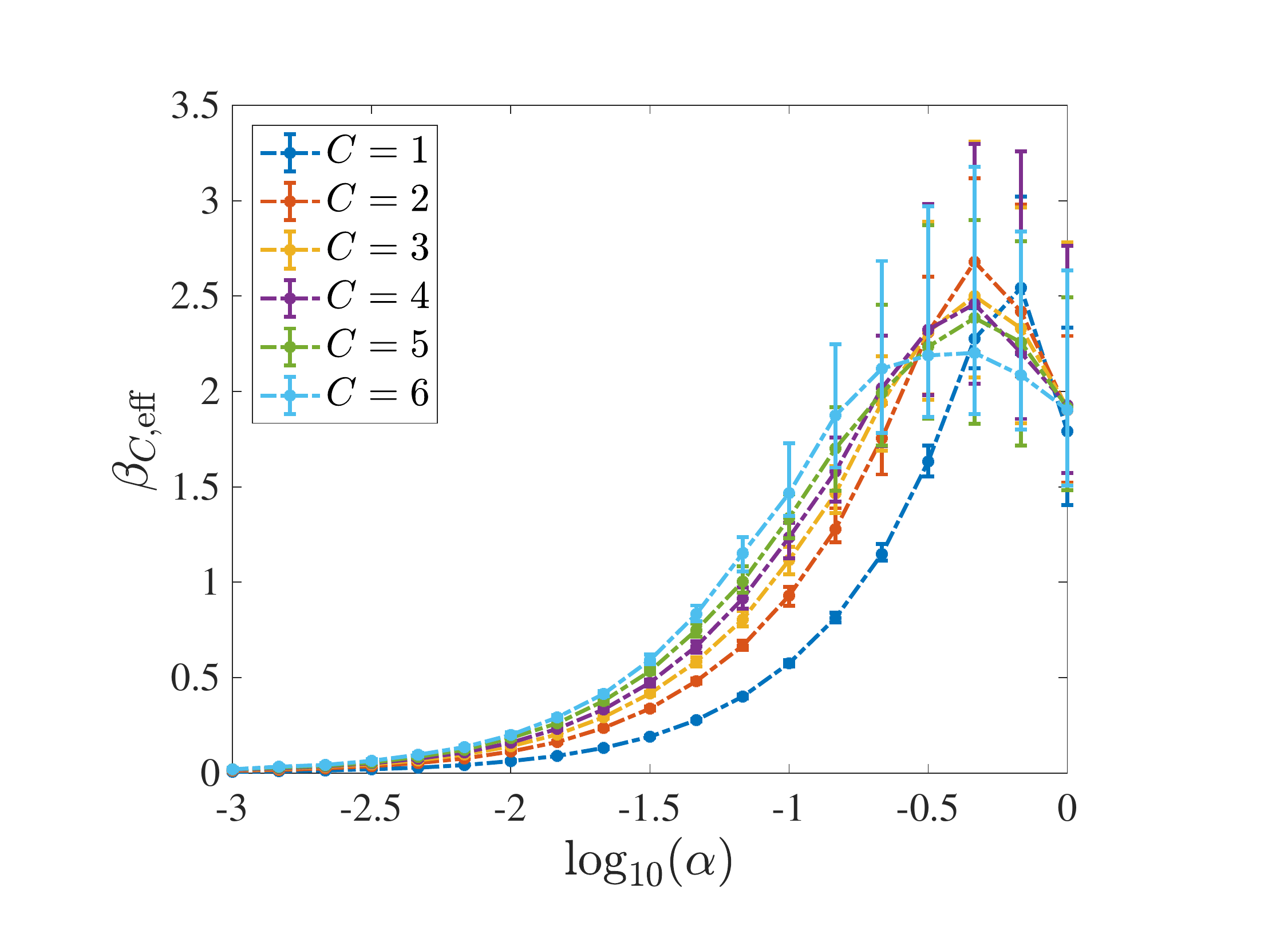}\label{fig:3a}}\hspace{-.6cm}
\subfigure[\,$K_{8}$: $\beta_{C,\rm eff}$, scaled]{\includegraphics[width=0.33\textwidth]{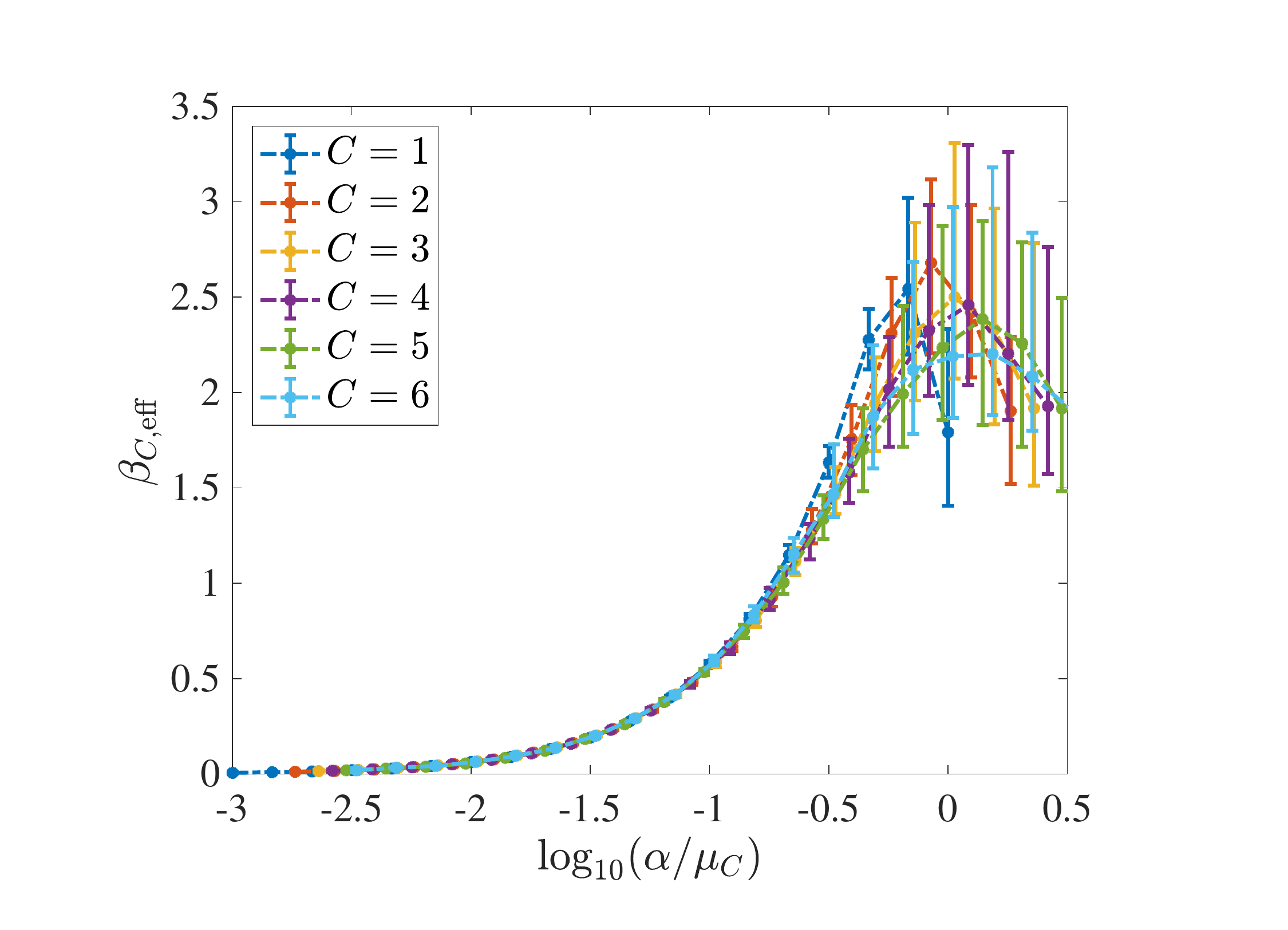}\label{fig:3b}}\hspace{-.6cm}
\subfigure[\,$K_{8}$: gradient overlap]{\includegraphics[width=0.33\textwidth]{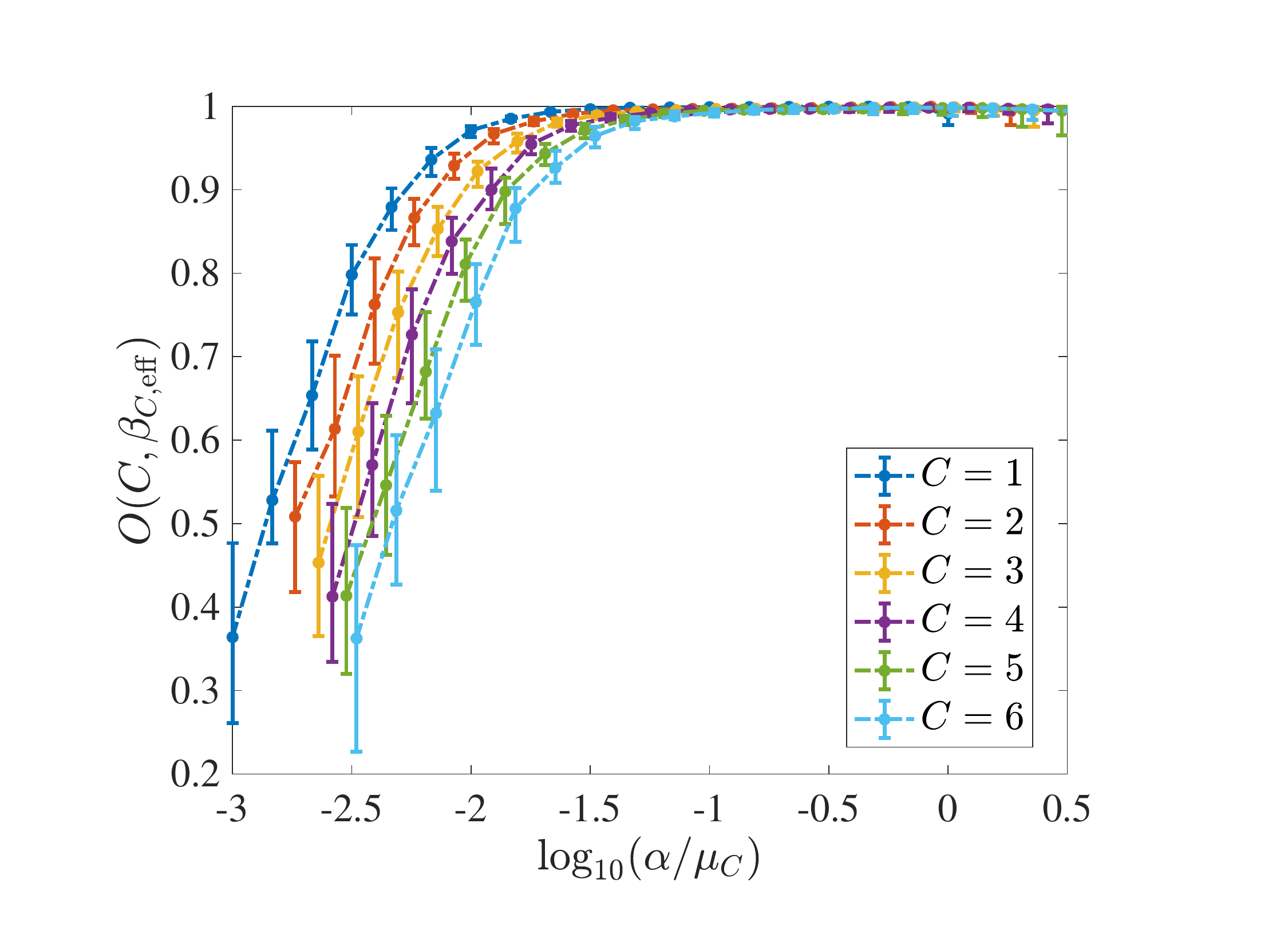}\label{fig:3d}}\\
\subfigure[\,$K_{16}$: $\beta_{C,\rm eff}$, unscaled]{\includegraphics[width=0.33\textwidth]{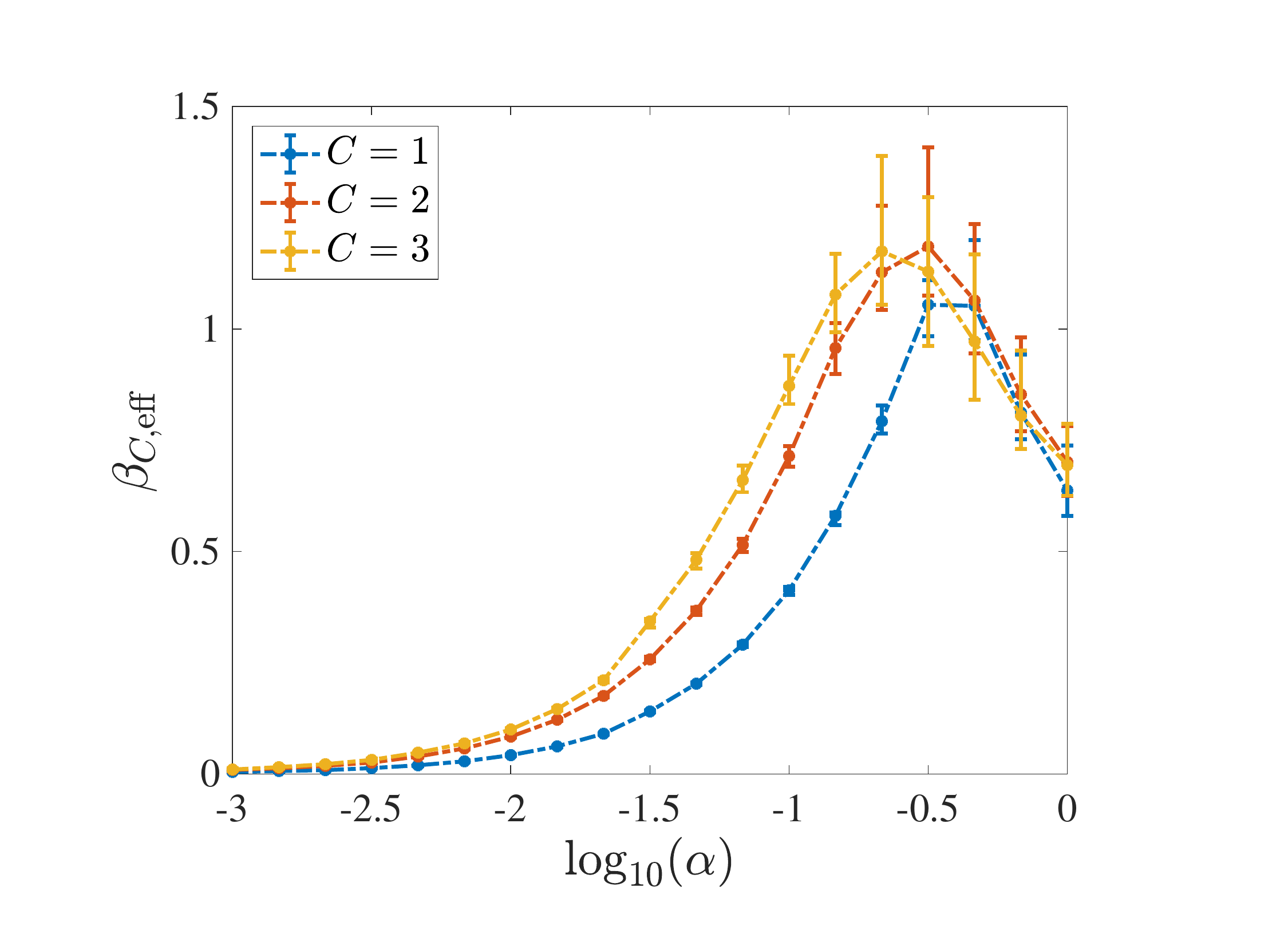}\label{fig:3e}}\hspace{-.6cm}
\subfigure[\,$K_{16}$: $\beta_{C,\rm eff}$, scaled]{\includegraphics[width=0.33\textwidth]{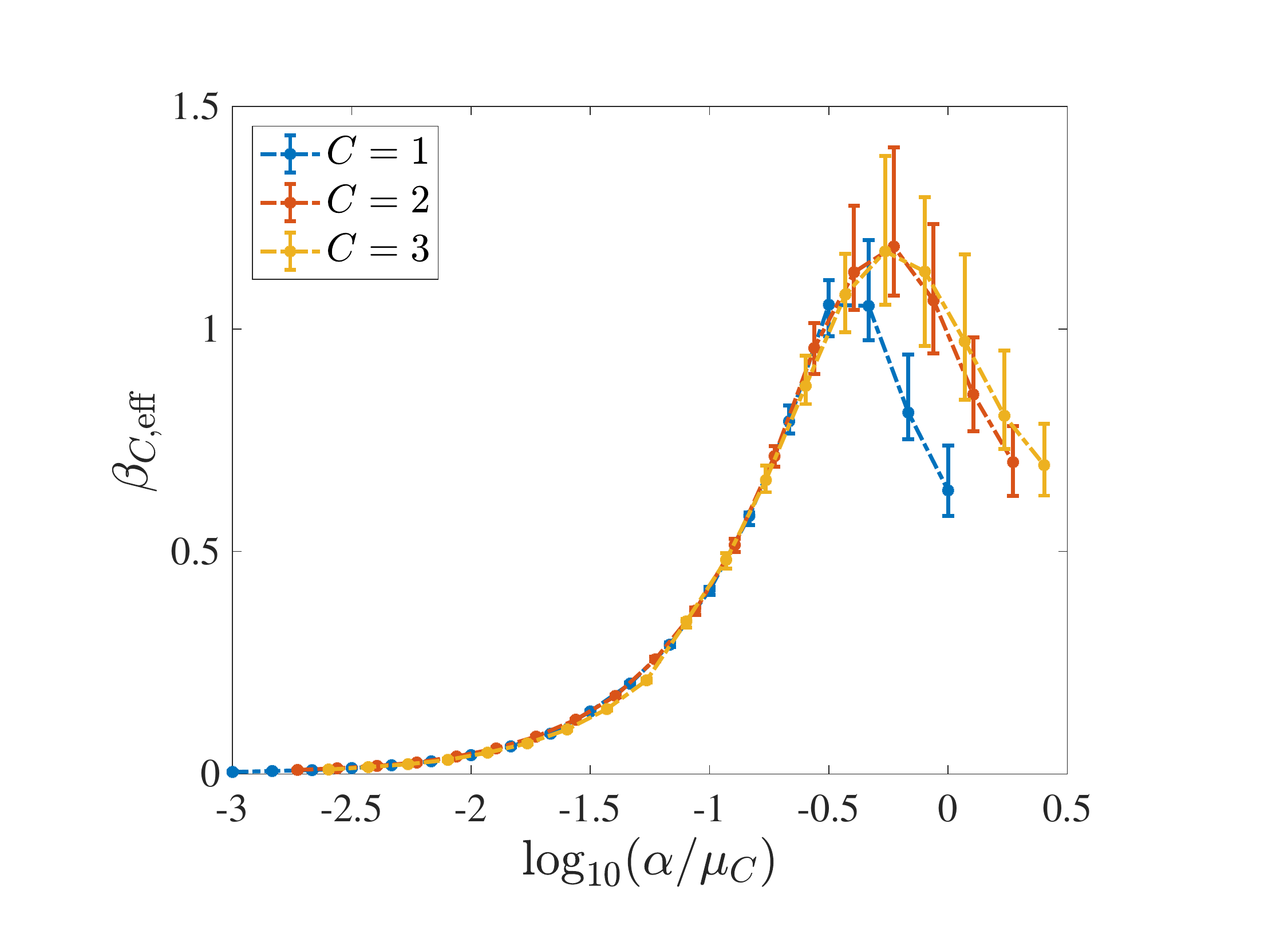}\label{fig:3f}}\hspace{-.6cm}
\subfigure[\,$K_{16}$: gradient overlap]{\includegraphics[width=0.33\textwidth]{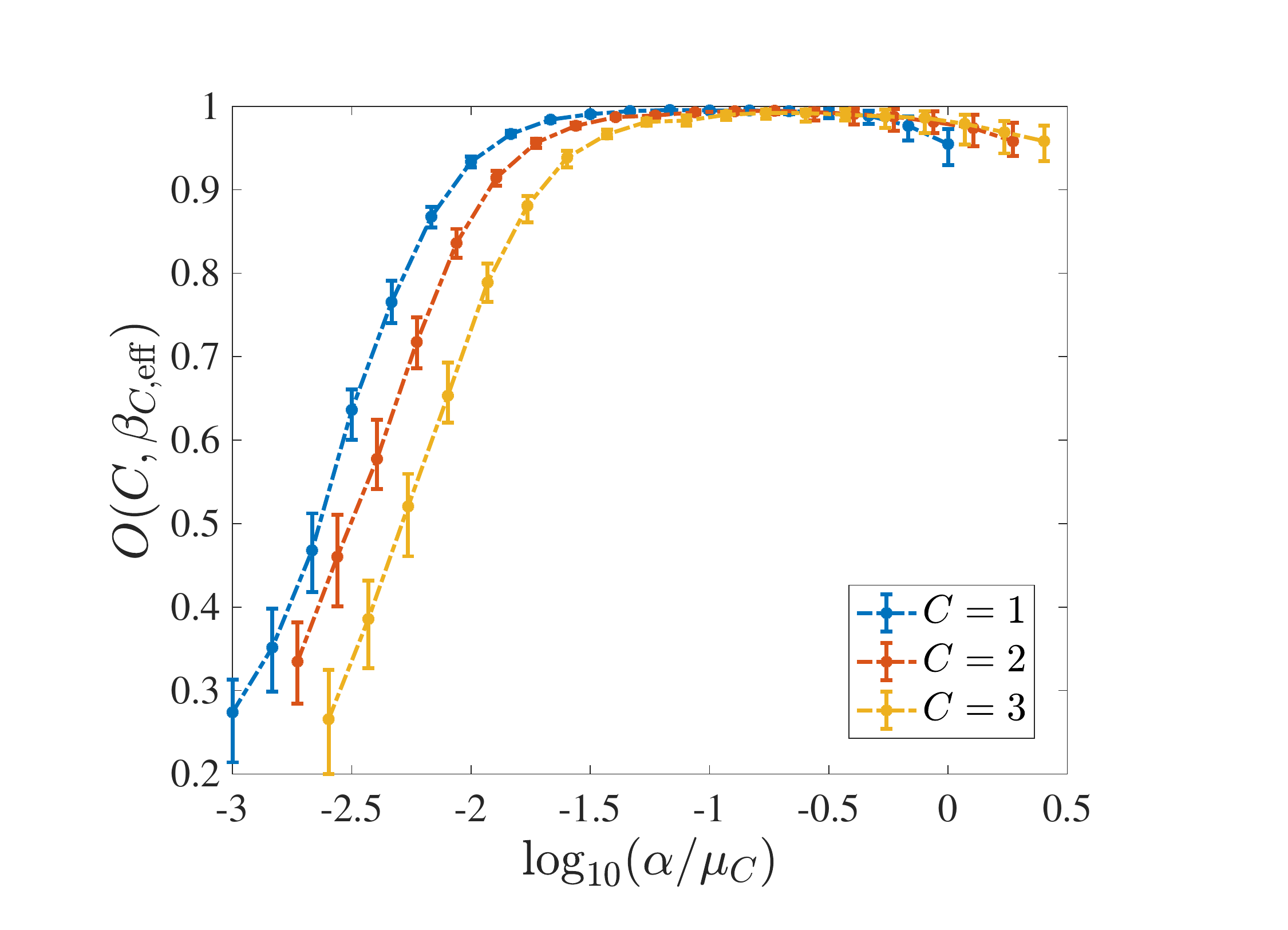}\label{fig:3h}}\\
\subfigure[\,$K_{24}$: $\beta_{C,\rm eff}$, unscaled]{\includegraphics[width=0.33\textwidth]{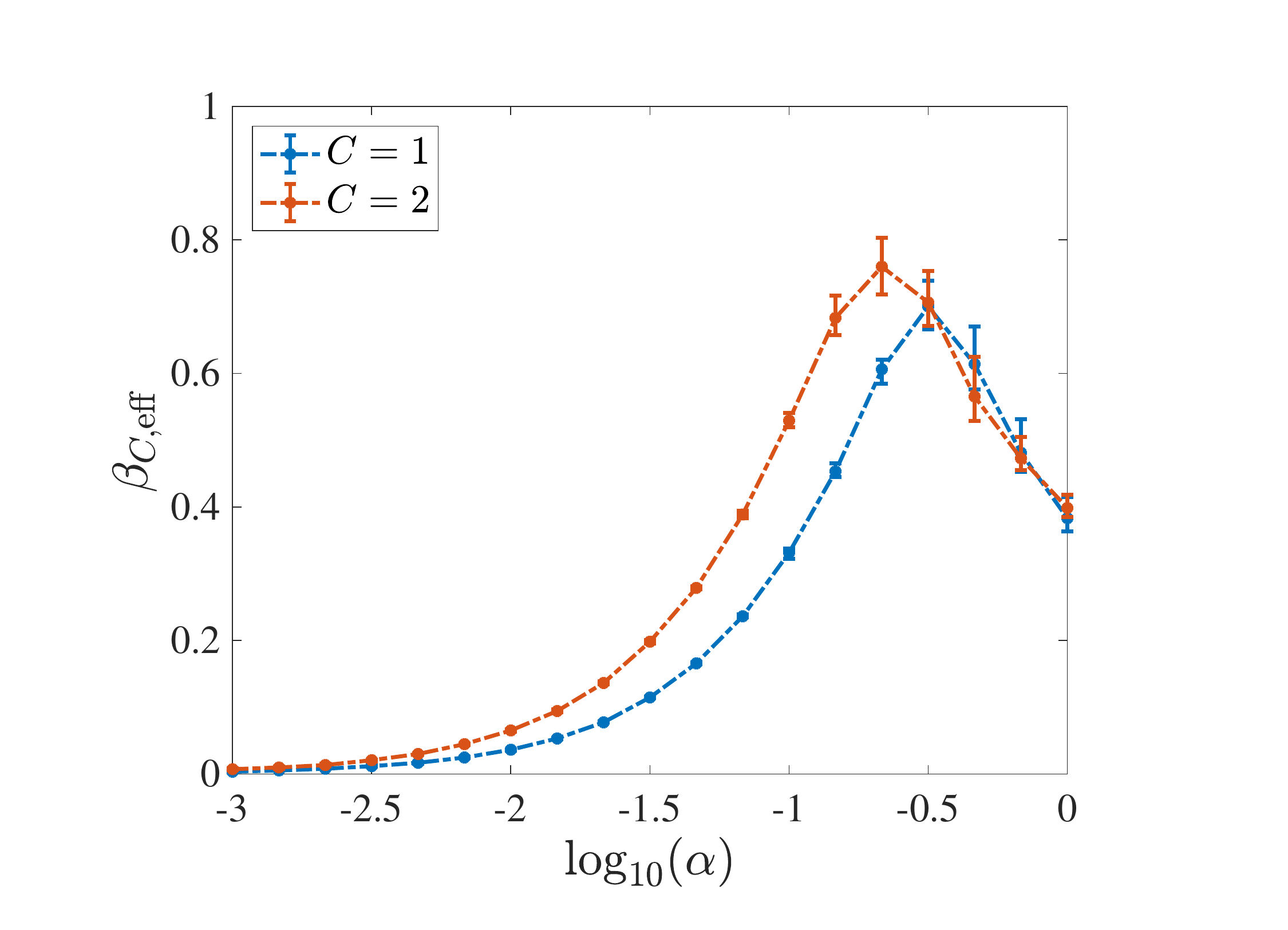}\label{fig:3i}}\hspace{-.6cm}
\subfigure[\,$K_{24}$: $\beta_{C,\rm eff}$, scaled]{\includegraphics[width=0.33\textwidth]{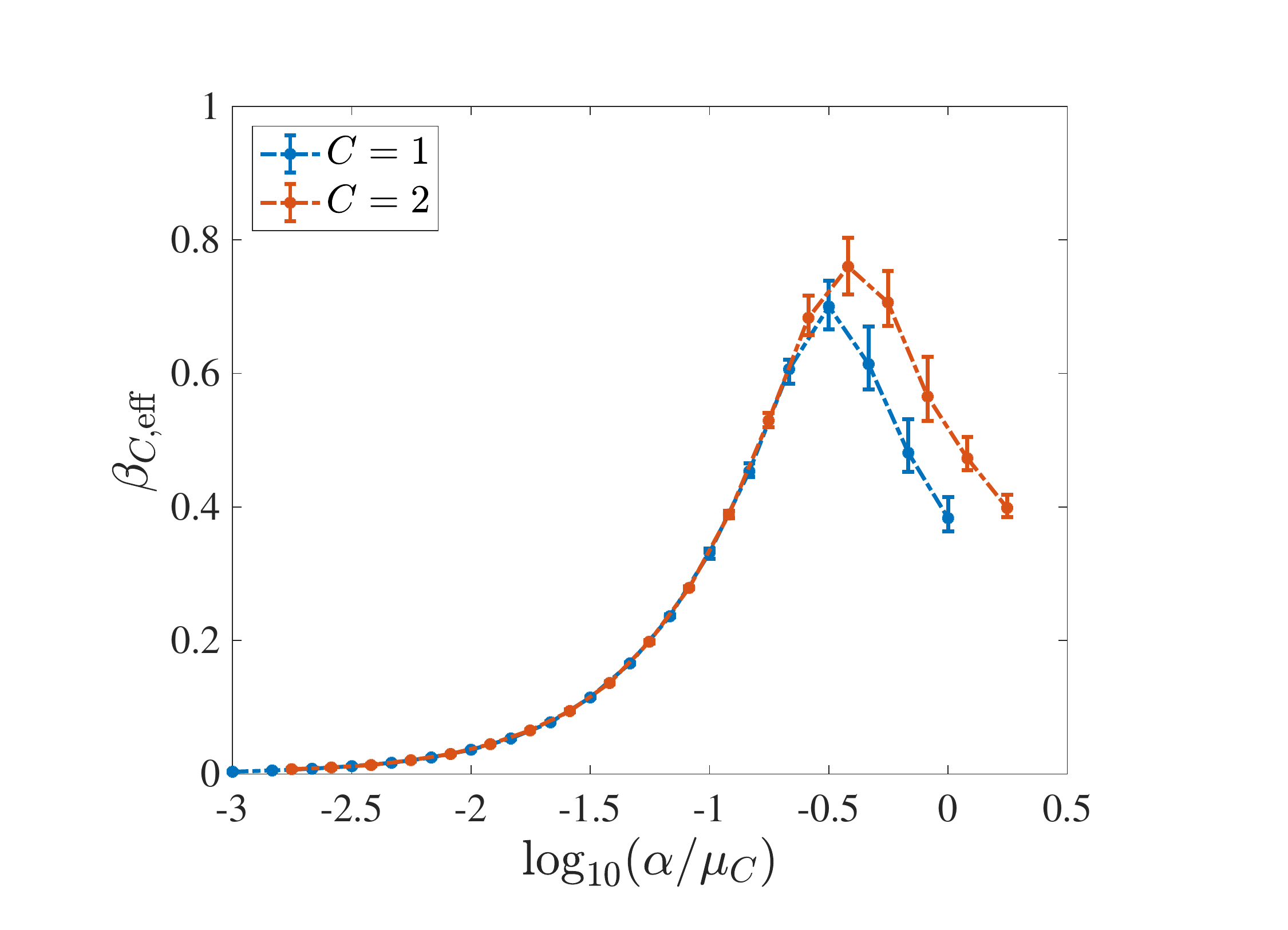}\label{fig:3j}}\hspace{-.6cm}
\subfigure[\,$K_{24}$: gradient overlap]{\includegraphics[width=0.33\textwidth]{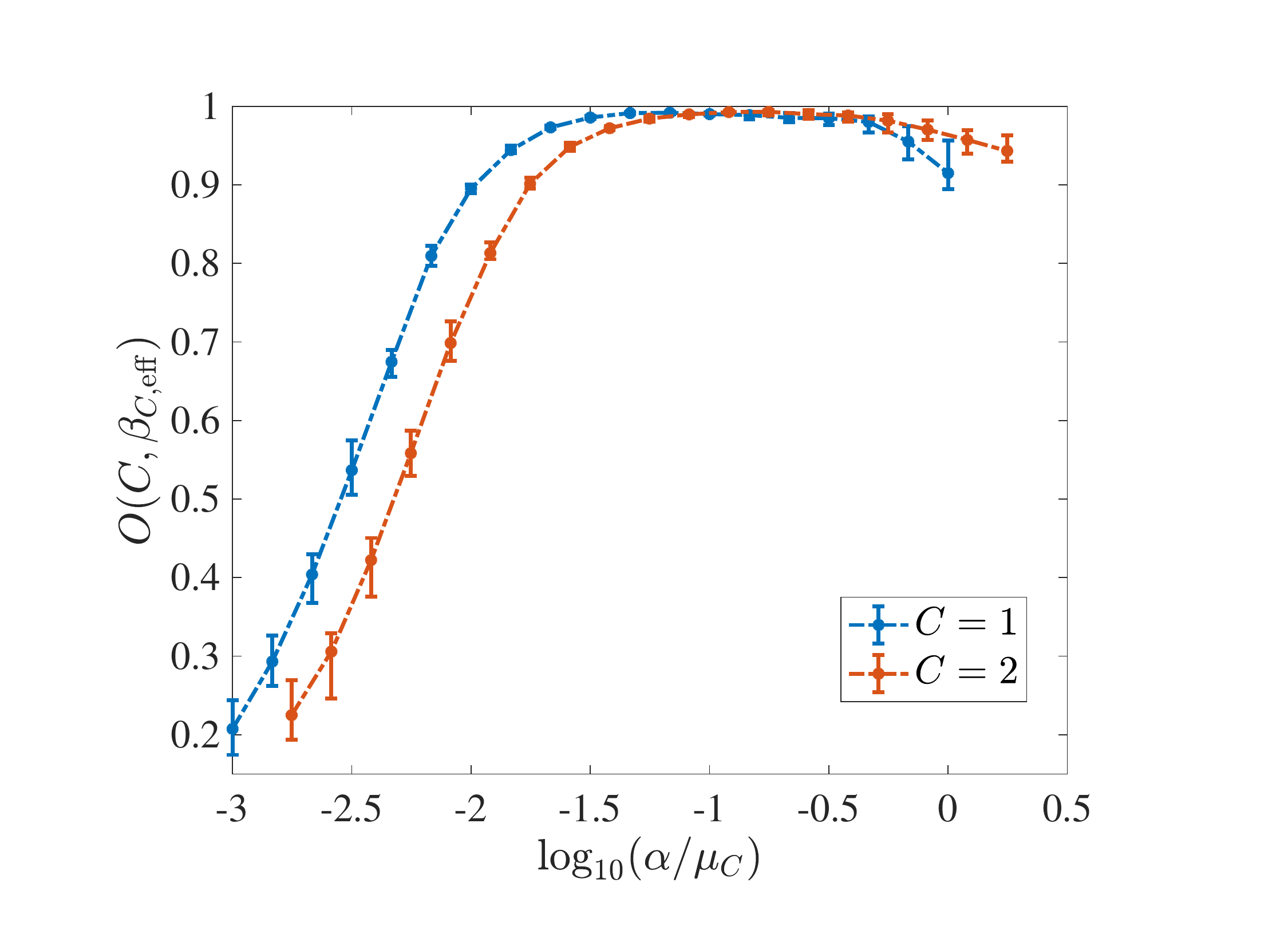}\label{fig:3l}}
\caption{Results for the $K_{8}$ ensemble (top row), $K_{16}$ ensemble (center row) and the $K_{24}$ ensemble (bottom row), of $100$ instances each. Leftmost panels show the effective inverse temperature $\beta_{C,{\rm eff}}$. Center
panels show the collapsed inverse temperature $\beta_{C,{\rm eff}}$.
Rightmost panels show the gradient overlap $O (p_{\rm DW}, p_T)$. In all plots we display the median values with the error bars ($25$th and $75$th percentiles) showing the variation over the ensembles. 
} 
\label{fig:3}
\end{center}
\end{figure*}

\subsubsection{The effective temperature associated with nesting level $C$}
We associate an effective inverse temperature with each nesting level $C$, which we denote by $\beta_{C, {\rm eff}}$. To find this quantity, consider the ``energy histogram":
\beq
p(E_a) =   \sum_{z | E(z)=E_a}  P(z)\,.
\label{eq:p(Ea)}
\eeq
This is the probability $p(E_a)$ to obtain a state with given energy $E_a$.

For each logical instance defined on a $K_N$, characterized by a set of logical local fields and couplings $\{h_i,J_{ij}\}$, we use Eq.~\eqref{eq:bi-wij} to define $b_i$ and $w_{ij}$, and numerically compute via Eq.~\eqref{eq:p(Ea)} the thermal energy histogram $p_T(E_a,\beta)$ [shorthand for $p_T(E_a,b_i= \beta h_i, w_{ij}= \beta J_{ij})$]. 
To do so we first compute $E(z)$ [Eq.~\eqref{eq:BE}] and 
the Boltzmann distribution $P(z)$ [Eq.~\eqref{eq:BP}].
Similarly, for the same logical instances we evaluate the corresponding empirical DW2KQ energy histogram $p_{\rm DW}(E_a,C)$ [shorthand for $p_{\rm DW}(E_a,C,h_i, J_{ij}) $], after the logical problem is encoded with $C$ nesting levels and the physical qubit readouts have been decoded via majority vote. 

The effective inverse temperature $\beta_{C,{\rm eff}}$ is now obtained by minimizing the total variation distance:
\bes
\begin{align}
\beta_{C,{\rm eff}} &= \arg \min_\beta D(p_{\rm DW}, p_T) \\
&=\arg \min_\beta\left(\frac12\sum_a |p_{\rm DW}(E_a,C) - p_T(E_a,\beta)|\right) \ .
\label{eq:dist}
\end{align}
\ees
In other words, the effective sampling temperature $1/\beta_{C,{\rm eff}}$ for the quantum annealer is defined as the temperature that minimizes the distance between the empirical and theoretical energy histograms $p_{\rm DW}(E,C)$ and  $p_T(E,\beta)$, respectively.\footnote{The total variation distance $D(p,q) \equiv \frac{1}{2}\sum_a |p_a - q_a|$ between the probability distributions $p$ and $q$ provides a simple way to estimate the effective sampling temperature which is good enough for our purposes. For more sophisticated approaches,  see for example Ref.~\cite{raymond2016global}.} Note that the theoretical energy histograms involve computing the partition function [Eq.~\eqref{eq:BP}] and so become numerically very demanding as $N$ grows. Indeed, this is precisely the reason that we are interested in the quantum annealing alternative.

\subsubsection{Assessing the quality of the sampling distributions obtained with NQAC}

The effective sampling temperature is an indirect measure of the quality of the sampling distribution obtained by the annealer. We thus consider an additional quantity that is more directly connected to the training of Boltzmann machines. Recall that the quantum annealer is used to evaluate the negative phases [Eq.~\eqref{eq:grads}] via sampling, i.e.,
\beq
\langle z_i z_j \rangle_T \equiv  \vec \nabla^T \mapsto  \langle z_i z_j \rangle_{\rm DW} \equiv  \vec \nabla^{\rm DW}  \,,
\eeq
where we have not included the quantities $\langle  z_i \rangle_{\rm DW}$ and $ \langle  z_i \rangle_{T}$ since for all the instances we have considered we have vanishing local fields, or biases, $h_{i}$. We regard the negative phases as vectors in the space of weights. This makes sense since the negative phases enter the definition of a ``gradient vector" as in Eq.~\eqref{eq:grads}. We thus consider the overlap between the thermal and experimental gradients:
\begin{align}
O (C,\beta_{C,{\rm eff}} ) &\equiv    \hat \nabla^{\rm DW}(C) \cdot \hat \nabla^T(\beta_{C,{\rm eff}})
\label{eq:quality}
\end{align}
where we have included a reminder that $O$ compares the empirical phases estimated with an NQAC encoding at nesting level $C$ and the negative phases numerically computed at temperature  $\beta = \beta_{C,{\rm eff}}$. 

\subsection{Empirical results}

\subsubsection{The effective temperature associated with nesting level $C$}

\begin{figure}[t]
\begin{center}
{\includegraphics[width=\columnwidth]{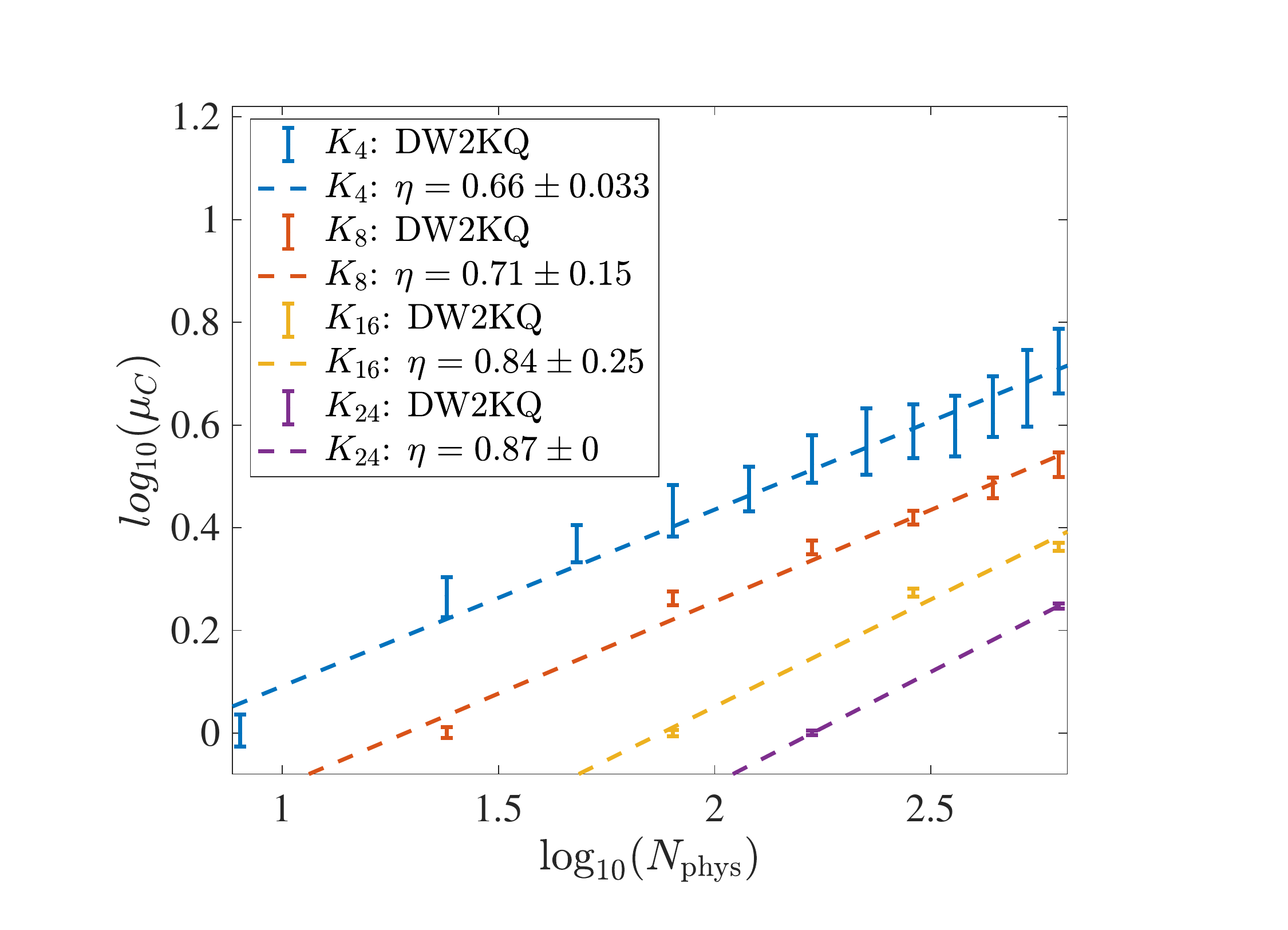}}
\caption{Scaling of the inverse effective temperature  reduction coefficient $\mu_{C}$ as a function of $C$ for the four ensembles of instances studied.  $\mu_{C}$ was computed as the rescaling coefficient that gives the data collapse in the center panels of Fig.~\ref{fig:3}. Empirical data are consistent with a power-law scaling $\mu_C \sim C^\eta$, where $\eta(K_4) = 0.66\pm0.033$\, $\eta(K_8) = 0.71\pm0.15$, $\eta(K_{16}) = 0.84\pm0.25$, $\eta(K_{24}) = 0.87\pm0$. See Appendix~\ref{app:B} for more details on the determination of $\mu_C$ and the associated error bars.} 
\label{fig:4}
\end{center}
\end{figure}

In Fig.~\ref{fig:3} we show the empirical results for the ensemble of instances $K_{8}$ [\ref{fig:3a}-\ref{fig:3d}],   $K_{16}$ [\ref{fig:3e}-\ref{fig:3h}] and $K_{24}$  [\ref{fig:3i}-\ref{fig:3l}]. Figures in the leftmost panels show the effective temperature $\beta_{C,{\rm eff}}$ computed via Eq.~\eqref{eq:dist}.
We observe a behavior similar to that of $P_C$ [see, e.g., Fig.~\ref{fig:1a}]. In the scaling region, $\beta_{C,{\rm eff}}$ grows monotonically with respect to both $\alpha$ and $C$. This shows that NQAC provides a systematic reduction of the effective sampling temperature. This interpretation is further confirmed in the center-left panels, in which we show a data collapse similar to that of Fig.~\ref{fig:1b}. In the scaling region we can find an energy boost $\mu_C$ such that
\beq
\beta_{C,{\rm eff}}(\alpha) \approx \beta_{1,{\rm eff}}(\mu_C \alpha)\ ,
\label{eq:mu_C-def2}
\eeq 
in analogy to Eq.~\eqref{eq:mu_C-def} for $P_C$. Thus, in the scaling region the effective sampling temperature of the annealer can be decreased at a given energy scale $\alpha$ of the problem Hamiltonian by increasing the nesting level $C$ [the scaling of $\mu_C$ is shown in Fig.~\ref{fig:4}].

\subsubsection{Assessing the quality of the sampling distributions obtained with NQAC}
 
The rightmost panels of Fig.~\ref{fig:3} show the gradient overlap $O(p_{\rm DW}, p_T)$.
We see that the numerically and empirically computed gradients $\vec \nabla$ are almost completely aligned for sufficiently large values of $\alpha$, in the scaling region.  At small values of $\alpha$ the gradient overlap shows a decay in sampling quality. Thus, there is a sweet-spot for using NQAC in sampling applications. For the examples shown here, this in the range $-1.5 \lesssim \log_{10}(\alpha) \lesssim -0.5$. In this range the empirically computed gradients are indistinguishable from the numerically computed one [as measured by $O (C,\beta_{C,{\rm eff}})$], 
and do not decay significantly with increased nesting level $C$; moreover, NQAC provides the added benefit of a monotonic reduction of the sampling temperature. 

Note that at small $\alpha$ (outside the ``sweet-spot region") we find that  $O (C,\beta_{C,{\rm eff}})$ is smaller for large $C$, so that higher nesting appears to be less helpful. However, recall that we did not optimize the energy penalty $\gamma$ (we set $\gamma = 1$), and in the small $\alpha$ region we expect that we can improve the sampling quality for larger $C$ by optimizing $\gamma$. Contrariwise, even without optimizing the energy penalty strength nesting improves matters for $\alpha$ larger than the sweet-spot region (this effect is only visible for the $K_{16}$ and $K_{24}$ cases, since for $K_8$ the overlap is pegged at $1$ for large $\alpha$).  

\subsubsection{Scaling of the energy boost}
In Fig.~\ref{fig:4} we show the scaling of $\mu_C$ as a function of the nesting level $C$ given by the data-collapse of the center panels of Fig.~\ref{fig:3}. For the $K_4$ case, Fig.~\ref{fig:4} shows a power-law scaling consistent with that we observed for the fully antiferromagnetic $K_4$, where we found $\eta = 0.68\pm0.06$ [Fig.~\ref{fig:1c}]. Interestingly, the result progressively improves for the larger graphs, with $K_{24}$ having the largest value of $\eta$, though this is admittedly tenuous as it is based on only only two data points due to the size limitations imposed by the DW2KQ. With this caveat in mind, we conclude that NQAC provides a lower effective sampling temperature scaling that improves with problem size. 

\ignore{
\subsubsection{Maximal achievable temperature reduction}
Rather than consider the scaling of the effective temperature over the entire range of $\alpha$ values, here we study the maximal temperature reduction achievable with NQAC. Considering  $\beta_{C,{\rm eff}}(\alpha)$ as a function of the energy scale $\alpha$, we define the smallest sampling temperature achievable at a given nesting level $C$ as   $\beta_{C,{\rm max}} =  \max_\alpha \beta_{C,{\rm eff}}(\alpha)$. We then consider the smallest achievable temperature among all implementable nesting levels: $\beta_{\rm max} = \max_C \beta_{C,{\rm max}}$. The ratio $\beta_{\rm max}/\beta_{1,{\rm max}}$ is a measure of the maximal temperature reduction that can be obtained via NQAC after optimizing the energy scale $\alpha$. In Fig.~\ref{fig:5} we show the results for the four ensembles of instances. The boxes in the figure represent the $25$th and $75$th percentiles (the red line is the median ratio) and the red crosses represent outliers. Figure~\ref{fig:5} shows that NQAC achieves a median maximal temperature reduction of about $10\%$. There are several outlying instances for which the temperature reduction is much more substantial (close to 2 times in some cases).
\begin{figure}[t]
\begin{center}
{\includegraphics[width=\columnwidth]{fig4b}}
\caption{Box-plot representation \cite{Frigge:1989vn} of the maximal temperature reduction achievable with NQAC.}
 \label{fig:5}
\end{center}
\end{figure}
}

\section{Discussion}
\label{sec:DISC}

Very recently it was pointed out that fixed finite temperature quantum annealers satisfy an adversarial ``temperature scaling law"~\cite{Albash:2017ab}, that prevents them from functioning as competitive scalable optimizers unless annealer temperatures are appropriately scaled down with problem size. This result places on a more rigorous footing the folklore wisdom that, in a realistic open system setting, AQC and QA require the temperature to be reduced as problem sizes grow and (generically) gaps shrink, since at a fixed temperature thermal transitions out of the ground state would become inevitable. Because the third law of thermodynamics (Nernst's ``unattainability principle" formulation, though see Ref.~\cite{Kolar:2012aa}) forces the cooling rate to vanish as the temperature approaches absolute zero, a scalable temperature reduction is considered impractical, known colloquially as there being ``no scaling law for refrigerators".

However, these perspectives ignore the possibility of error correction, which acts as an effective entropy sink, and can be used to attain a scalable temperature reduction at the expense of using up physical qubits to create colder code qubits \cite{jordan2006error,Marvian:2017aa}. In this work we addressed and provided a potential solution for the ``temperature scaling law" problem, by demonstrating that NQAC allows for a scalable effective temperature reduction, by using codes of increasing code distance $C$ to achieve an effective temperature reduction scaling as a power law in $C$. This potentially addresses even the most adversarial scenario considered in Ref.~\cite{Albash:2017ab}, wherein the temperature must drop as a power law with problem size. 

Our work extends the results and analysis of Ref.~\cite{vinci2015nested}, which introduced NQAC, in two significant ways. 

First, we confirmed that NQAC provides a performance improvement that can be interpreted as an effective temperature reduction when the method is implemented on a DW2KQ quantum annealer (the DW2KQ device has four times as many physical qubits as the DW2 device used in Ref.~\cite{vinci2015nested}). This allowed us to consider NQAC encodings with larger nesting levels (or code distance) $C$; e.g., from $C=8$ on the DW2 device to $C=13$ in the present case, for $K_4$ problem instances. Moreover, we were able to study the encoding of larger problems (up to $C=2$ for $K_{24}$ problem instances). We confirmed the existence of a scaling law $T_{\rm eff} \sim C^{-\eta}$ that is valid in the scaling regime, i.e., when the energy scale factor $\alpha$ is small enough that the NQAC implementation is not limited by the strength of the energy penalties. In this scaling regime NQAC outperforms classical repetition, a benchmark for any bona fide error correcting scheme.

Second, for the first time we studied the use of NQAC in sampling applications, specifically estimating the gradient required for training Boltzmann machines. We argued that there are two important quantities that characterize the performance of quantum annealers as samplers: the sampling temperature and the quality of sampling. Improving the performance of a quantum annealing device for sampling via error correction should reduce the sampling temperature of the device without diminishing the quality of the sampling. We showed that NQAC can indeed achieve a monotonic reduction of the sampling temperature. As in the case of optimization applications, this holds in the scaling regime, i.e., when the encoding is not limited by the strength of the energy penalties used in the encoding. 

We do not expect the NQAC method studied here to provide an indefinite effective temperature reduction at arbitrary problem sizes; such a result belongs in the realm of the open problem of fault-tolerant AQC and QA. However, we do expect that the effective temperature reduction we have demonstrated here for both optimization and sampling applications to fuel the continued study of practical and implementable error suppression schemes for QA, and perhaps even carry the next generation of quantum annealers to the point of demonstrating a quantum advantage.

\acknowledgments
%
We thank D-Wave Systems Inc. for access to their DW2KQ device in Burnaby. This work was (partially) supported under ARO grant number W911NF-12-1-0523, ARO MURI Grant Nos. W911NF-11-1-0268 and W911NF-15-1-0582, and NSF grant number INSPIRE-1551064. The research is based upon work partially supported by the Office of the Director of National Intelligence (ODNI), Intelligence Advanced
Research Projects Activity (IARPA), via the U.S. Army Research Office
contract W911NF-17-C-0050. The views and conclusions contained herein are
those of the authors and should not be interpreted as necessarily
representing the official policies or endorsements, either expressed or
implied, of the ODNI, IARPA, or the U.S. Government. The U.S. Government
is authorized to reproduce and distribute reprints for Governmental
purposes notwithstanding any copyright annotation thereon.

\appendix

\section{Experimental details}
\label{app:A}

\begin{figure*} %
   \centering
\subfigure[]{\includegraphics[width=0.95\columnwidth]{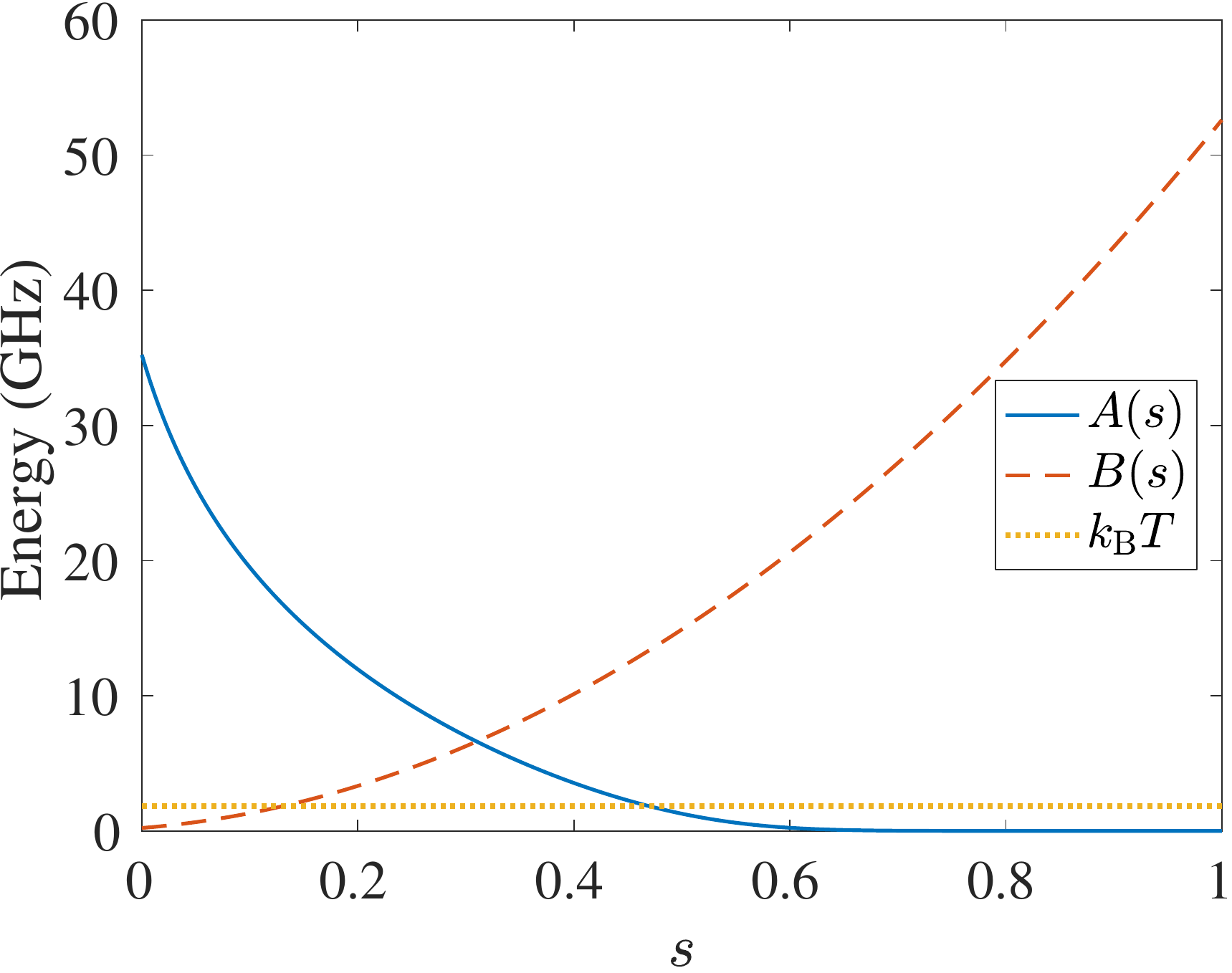}\label{fig:DW2000Qannealing}}
\hspace{.5cm}
\subfigure[]{\includegraphics[width=0.95\columnwidth]{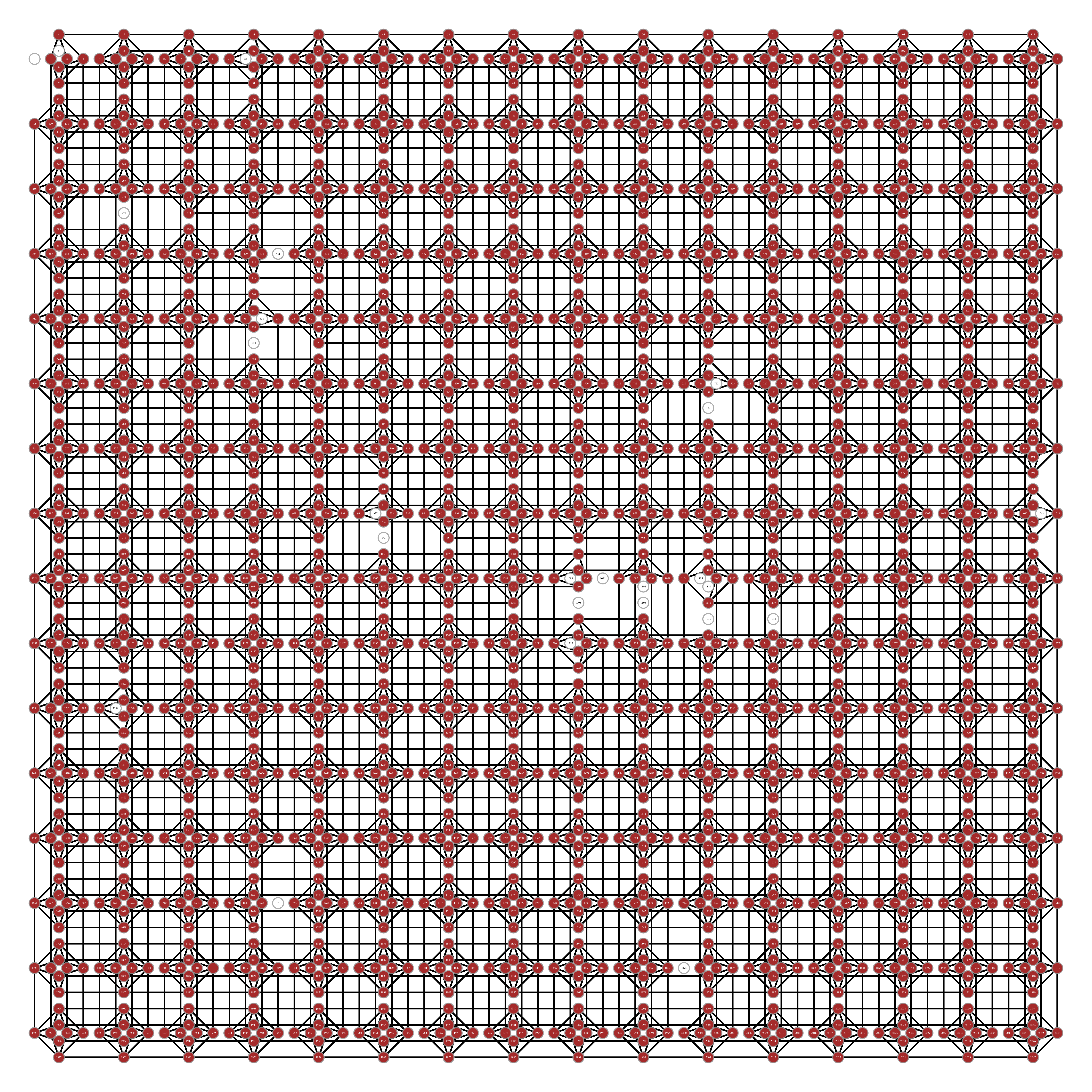}\label{fig:hardwaregraph}}
   \caption{(a) Annealing schedule for the DW2KQ (with $\hbar = 1$ units).  As a reference, we include the operating temperatures of the device, corresponding to $14.1$mK for the DW2KQ. (b) ``Chimera" hardware graph of the DW2KQ device in Burnaby used in this work. Available qubits are shown in green, and unavailable qubits are shown in red.  Programmable couplers are shown as black lines connecting qubits.}
   \label{fig:AnnealingSchedules}
\end{figure*}

The annealing schedules of the D-Wave 2000Q (DW2KQ) processor housed in Burnaby is shown in Fig.~\ref{fig:DW2000Qannealing}. 
The Chimera hardware graphs of the DW2KQ processor we used in this work  is shown in Fig.~\ref{fig:hardwaregraph}. For additional experimental details see, e.g., Refs.~\cite{Harris:2010kx,Bunyk:2014hb,Albash:2017aa}.

\section{Determination of the energy boost $\mu_C$ and error bars}
\label{app:B}

Figures~\ref{fig:1c} and \ref{fig:4} show the experimental scaling of $\mu_C$ as extracted by using, respectively, Eq.~\eqref{eq:mu_C-def} (via success probability) and Eq.~\eqref{eq:mu_C-def2} (via effective temperature). To determine the values of $\mu_C$ and estimate error bars, we proceeded as follows. Let us define $M_C(\alpha) \equiv P_C(\alpha)$ or $M_C(\alpha) \equiv \beta_{C,{\rm eff}}(\alpha)$ depending on whether we are using $P_C(\alpha)$ [Fig.~\ref{fig:1c}] or $\beta_{C,{\rm eff}}(\alpha)$ (Fig.~\ref{fig:4}).

First, we used smoothing splines to determine a continuous  interpolation $M^{\mathrm{mid}}_C(\alpha)$ of the median data points of $M_C(\alpha)$. In the same way we also determined the higher and lower interpolating curves $M^{\mathrm{high}}_C(\alpha)$ and $M^{\mathrm{low}}_C(\alpha)$ for the data points of the $75$-th and $25$-th percentiles of $M_C(\alpha)$ respectively. Then,  a reference value $\alpha^{\mathrm{mid}}_C$ was determined such that $M_C^{\mathrm{mid}}(\alpha^{\mathrm{mid}}_C) = M_0$, where we used the smooth interpolation of the experimental data. The energy boost was then determined as $\mu_C =  \alpha^{\mathrm{mid}}_1/\alpha^{\mathrm{mid}}_C$.  $M_0$ is an arbitrarily chosen reference value where the different $M_C(\alpha)$ curves are made to overlap. This reference  serves as a base point for computing $\mu_C$. As shown in the center panels of Fig.~\ref{fig:1} and~\ref{fig:3}, the overlap of the $P_C$ and  $\beta_{C,{\rm eff}}(\alpha)$ data over the entire $\alpha$ range means that the specific choice of $M_0$ is arbitrary.  We similarly determined  $\mu^{\mathrm{high}}_C =  \alpha^{\mathrm{high}}_1/\alpha^{\mathrm{high}}_C$ and $\mu^{\mathrm{low}}_C =  \alpha^{\mathrm{low}}_1/\alpha^{\mathrm{low}}_C$ using the corresponding interpolating curves. The error bars shown in the figures were then centered at  $\mu_C$, with lower and upper error bars being $\mu^{\mathrm{high}}_C$ and $\mu^{\mathrm{low}}_C$, respectively.

%
%

\bibliography{refs}

\end{document}